\documentclass[journal]{IEEEtran}
\usepackage{ifpdf}
\RequirePackage{fix-cm}
\usepackage{color}
\usepackage[dvipsnames]{xcolor}
\usepackage{cite}
\usepackage{hyperref}
\usepackage{pifont}
\hypersetup{
    hidelinks 
}
\usepackage{bbm}
\usepackage[caption=false,font=footnotesize]{subfig}
\usepackage{tikz}

\newcommand{\circlednum}[1]{%
  \tikz[baseline=(char.base)]{
    \node[
      draw,
      circle,
      inner sep=0.4pt,
      line width=0.4pt
    ] (char) {\scriptsize #1};
  }%
}
\usepackage{graphicx}
\ifCLASSINFOpdf
\else
\fi
\usepackage{amsmath}
\usepackage{algpseudocode}
\usepackage{array}
\usepackage{tabularx}
\usepackage{booktabs}
\usepackage{multirow}
\usepackage{enumitem}
\usepackage{threeparttable}
\usepackage{amsmath,amssymb,bm}
\usepackage[table]{xcolor}
\usepackage[ruled,vlined,linesnumbered]{algorithm2e}
\SetAlgoNlRelativeSize{-1}
\usepackage{url}
\usepackage{float} 
\begin{document}

\title{\huge Cross-View Vision-Aided Proactive BS Selection and Beam Prediction for mmWave V2I Communications}

\author{Zijiao Hu,
Haiyao Yu,
Gaoyang Pang,
Guangchen Wang,
Litianyi Zhang,\\
Wanchun Liu,~\IEEEmembership{Senior Member,~IEEE},
George C. Alexandropoulos,~\IEEEmembership{Senior Member,~IEEE},\\
Branka Vucetic,~\IEEEmembership{Life Fellow,~IEEE},
and Yonghui Li,~\IEEEmembership{Fellow,~IEEE}

\vspace{-0.6cm}

\thanks{

Z. Hu, H. Yu, G. Pang, G. Wang, L. Zhang, W. Liu, B. Vucetic, and Y. Li are with the School of Electrical and Computer Engineering, University of Sydney, Sydney, NSW 2006, Australia (e-mails: \{zijiao.hu, haiyao.yu, gaoyang.pang, guangchen.wang, litianyi.zhang, wanchun.liu, branka.vucetic, yonghui.li\}@sydney.edu.au).

G. Wang is also with the School of Engineering, Australian National University,
Canberra, ACT 2600, Australia (e-mail: guangchen.wang@anu.edu.au).

G. C. Alexandropoulos is with the Department of Informatics and
Telecommunications, National and Kapodistrian University of Athens,
Athens 16122, Greece (e-mail: alexandg@di.uoa.gr).
}
}


\maketitle
\begin{abstract}
This paper investigates {environmental}-sensing-aided proactive {base station (BS)} selection and beam prediction for millimeter-wave (mmWave) vehicle-to-infrastructure (V2I) {wireless} systems. We exploit onboard panoramic street-view images and a preloaded satellite map to predict communication-relevant environmental information around the vehicle, including nearby building footprints and heights. The predicted height map provides a compact environmental prior and is combined with historical mobility information to jointly predict the next-slot line-of-sight (LoS) state, transmission rate, and transmit and receive beam selections. On our dataset covering different real-world regions across New South Wales, Australia, the proposed framework achieves $\textbf{91.4\%}$ LoS classification accuracy, $\textbf{0.638}$~bps/Hz mean absolute error of data rate prediction, and more than 40\% higher transmission rate than the conventional reactive baseline in geographically unseen regions, outperforming all evaluated deployable learning-based baselines. The dataset and code {will be released} at \href{https://github.com/Huzijiao/Cross-view_V2I}{https://github.com/Huzijiao/Cross-view\_V2I}.
\end{abstract}


\IEEEpeerreviewmaketitle

\section{Introduction}

\IEEEPARstart{W}{ith} the development of 6G and deployment of ultra-dense networks (UDNs), vehicle-to-infrastructure (V2I) communications are expected to support ultra-reliable, high-data-rate, and low-latency vehicular services. Millimeter-wave (mmWave) transmission is a key technology for the 6G V2I communications~\cite{noor6GV2X2022} because of its high carrier frequency and large available bandwidth, which can substantially improve link capacity in vehicular networks. However, the high-frequency propagation also makes mmWave links vulnerable to severe path loss, attenuation, and weak diffraction~\cite{baiCoverageRateAnalysis2015,rappaportOverviewMillimeterWave2017}. As a result, designing the beamforming matrices and handover strategies to achieve reliable mmWave V2I communications becomes complex and challenging, especially in high-mobility urban scenarios.

Conventional V2I beamforming and handover methods are mainly classified into two groups~\cite{giordaniTutorialBeamManagement2019}. The first group is beam sweeping over a predefined transceiver codebook~\cite{xiaoHierarchicalCodebookDesign2016,aykinEfficientBeamSweeping2020}, which sweeps the beam to select the beam pair or serving base station (BS) that maximizes the received power or transmission data rate. The second group relies on uplink pilots or reference signals from the user equipment (UE) to estimate the channel, infer the UE location or angle information~\cite{vaInverseMultipath2018,9685723}, and then design the beamforming matrix and handover strategy accordingly. Although these approaches are effective in quasi-static links, they become expensive in vehicular scenarios. Specifically, the exhaustive or hierarchical beam sweeping overhead grows with the codebook size and the number of candidate BSs~\cite{baratiInitialAccess2016,liDesignAnalysisInitialAccess2017,aykinEfficientBeamSweeping2020}. The pilot-based channel estimation also requires frequent measurement, feedback, and reconfiguration, while the estimated channel degrades rapidly when the vehicle moves at high speed~\cite{vaImpactBeamwidth2017} or when the link state changes suddenly~\cite{charanVisionAided2021}. Therefore, reactive beam design incurs substantial beam searching or training overhead and initiates handover only after the link quality has already degraded, leading to reduced connection stability, delayed handover, and less effective data transmission time.

Recently, integrated sensing and communication (ISAC) has attracted increasing attention for V2I beam management and link prediction~\cite{liuIntegratedSensing2022}. The main intuition is that sensing side information, such as global positioning system (GPS) location measurements~\cite{8313072, 10266792}, radar~\cite{luoMillimeterWaveV2V2023}, light detection and ranging (LiDAR) point clouds~\cite{jiangLiDARAidedFuture2023}, and RGB/depth images from cameras mounted at either the BS or the vehicular user (VUE)~\cite{charanVisionAided2021,xuComputerVisionAided2023}, can capture propagation-relevant environmental features, such as line-of-sight (LoS) paths, blockage, scatterers, and user--BS geometry. These features can guide the design of proactive mmWave beam management strategies that reduce beam-searching overhead and handover latency. For example, a LiDAR-aided learning framework has been proposed in~\cite{jiangLiDARAidedFuture2023} to exploit point-cloud data with GPS to predict both current and future beam indices in real-world mmWave V2I scenarios. Extending sensing-aided beam management to vehicle-to-vehicle (V2V) scenarios, a radar-aided V2V beam tracking framework has been proposed in~\cite{luoMillimeterWaveV2V2023} to use the range-{D}oppler map to reconstruct surrounding moving vehicles, which is utilized to predict the future beams between moving vehicles. 
However, LiDAR- or radar-based sensing typically incurs substantial hardware, calibration, and deployment costs. Although such sensors can provide accurate geometric information, their cost and infrastructure requirements limit their practicality for large-scale V2I deployment~\cite{zhengVisionAssisted2023}.

Vision-aided V2I communications have become feasible because cameras are low cost, easy to deploy, and capable of capturing rich environmental information without occupying additional wireless spectrum~\cite{charanVisionAided2021,10412143,yangEnvironmentSemantics2023}. Existing vision-aided methods can be broadly divided into BS-side and vehicle-side sensing~\cite{ahnSensingComputerVisionAided2024, liOutofBandModalitySynergyBased2026a}. In BS-side sensing, cameras mounted at BSs, roadside units (RSUs), or third-party infrastructures observe the road scene and assist beam alignment, blockage prediction, or link scheduling. Recently, object detection results and position embeddings from BS-mounted RGB camera images have been used for future link-rate prediction and link switching~\cite{ahnSensingComputerVisionAided2024}. Moreover, the authors in~\cite{liOutofBandModalitySynergyBased2026a} construct binary encoding maps (BEM) from BS-side RGB camera images and BS/vehicle location information by extracting bounding boxes of multiple users and scatterers, and use the resulting BEM sequences to predict user-specific channel gains, beam indices, and proactive BS selection. Nevertheless, BS-side camera deployment is difficult to scale in dense urban networks because each BS requires camera installation, calibration, maintenance, and field-of-view planning.  BS-side visual monitoring also introduces privacy concerns because road users and surrounding objects are continuously captured from infrastructure cameras~\cite{xuComputerVisionAided2023}. In contrast, vehicle-side vision can exploit sensing devices that already exist on the VUE, especially in autonomous-driving or advanced driver-assistance systems, where cameras are used for navigation, localization, and obstacle avoidance~\cite{xuComputerVisionAided2023}. Existing vehicle-side vision-aided works use images captured by the VUE to detect surrounding vehicles, estimate dynamic channel conditions, and predict beam alignment and handover strategies~\cite{osmanVehicleCamerasGuide2023,fabianiMultiModalSensingCommunication2024}.

However, most of the forediscussed studies focus on the blockage between nearby vehicles or between a roadside RSU and the target vehicle~\cite{luoMillimeterWaveV2V2023,osmanVehicleCamerasGuide2023,fabianiMultiModalSensingCommunication2024,jiangLiDARAidedFuture2023,xuComputerVisionAided2023}. In practical V2I deployments, when the antenna is mounted on the roof of the VUE and communicates with elevated BSs such as RSUs, rooftop BSs, or unmanned aerial vehicle (UAV)-mounted BSs, the dominant blockage is often induced by surrounding buildings rather than only by neighboring vehicles~\cite{begishevClosedFormUAVBlockage2022, wangMmWaveV2I2018 }. Existing works either do not explicitly consider the impact of building geometry or assume that building footprints and their heights are known from accurate maps~\cite{levieRadioUNet2021}. This assumption is restrictive because building heights are unavailable or incomplete in many cities, and urban maps may be incomplete or outdated as the environment evolves. Therefore, how to use vehicle-accessible visual observations to predict surrounding building shapes and heights for V2I beam and handover decisions remains insufficiently investigated.

In this paper, motivated by this gap, we propose a vision-aided proactive V2I prediction framework, which utilizes the onboard camera-captured street-view images and preloaded satellite map to capture the surrounding environmental information around the VUE, and then predicts the next-slot beam selection and handover decisions.
The main contributions of this paper are summarized as follows:
\begin{itemize}
\item To the best of our knowledge, this work is among the first to incorporate height-map estimation from cross-view visual observations into a communication-aware V2I decision-making pipeline. Specifically,
we propose a visual building height-map estimation (VBHE) model that utilizes both satellite and street-view images for building-footprint segmentation and height estimation. The estimated height map captures link propagation-relevant geometric information around the VUE, which is then used to guide the subsequent multi-task link prediction.

\item We propose a multi-modal multi-task spatial-temporal model (MMST) that extracts candidate-specific geometric features for each VUE--BS link from the predicted height map. These features are fused with the historical mobility information to jointly predict multiple link tasks, including the LoS probability, transmission rate, and transmit and receive beam indices. These predictions are then used to guide proactive BS selection and beam prediction for the next time slot along the VUE trajectory.
\item We construct a dataset covering 18 real-world urban regions across New South Wales, Australia, including the panoramic street-view images, bird's-eye-view (BEV)-domain satellite images, and the ground truth building-footprint and height maps. We then use the ray-tracing platform Sionna to generate the corresponding V2I channel labels along different trajectories of these regions. The resulting dataset contains $17$K visual samples for VBHE and ${290}$K communication-channel samples for MMST. Simulation results on this dataset show that both VBHE and MMST outperform the baselines in height-map estimation and multi-task link prediction, respectively, in geographically unseen test regions.
\end{itemize}

\textbf{Notations}: Uppercase (e.g., $X$) and lowercase (e.g., $x$) letters denote constants and scalar variables, respectively, while $\mathbb{R}$ is the set of real numbers. Sets are represented by calligraphic letters (e.g., $\mathcal{A}$), with cardinality $|\mathcal{A}|$ and element sequence explicitly denoted as $\{x_{\ell}\}_{\ell = a,\ldots,b}$. Bold uppercase (e.g., $\mathbf{A}$) and lowercase (e.g., $\mathbf{a}$) letters designate matrices or tensors, and column vectors, respectively. The superscript $^\top$ and $^{\mathrm{H}}$ indicate transposition and Hermitian transpose, respectively. We use $\mathbf{a}\! \sim \!\mathcal{CN}(\mu, \Sigma)$ to indicate that vector $\mathbf{a}$ is complex Gaussian distributed with mean $\mu$ and covariance matrix $\Sigma$. Given matrices $\mathbf{A}$ and $\mathbf{B}$, $[\mathbf{A},\mathbf{B}]$, $[\mathbf{A}^\top,\mathbf{B}^\top]^\top$, $\mathbf{A}\otimes\mathbf{B}$, and $\mathbf{A} \odot \mathbf{B}$ denote horizontal, vertical concatenations, the Kronecker product, and element-wise product, respectively. $\lfloor{x}\rfloor$ denotes the floor function, and $\mathbf{A}[{i,j}]$ denotes the $(i,j)$-th entry of the matrix $\mathbf{A}$.

\section{System Model and Problem Formulation}
\begin{figure}[t]
\centering
\includegraphics[width=\linewidth]{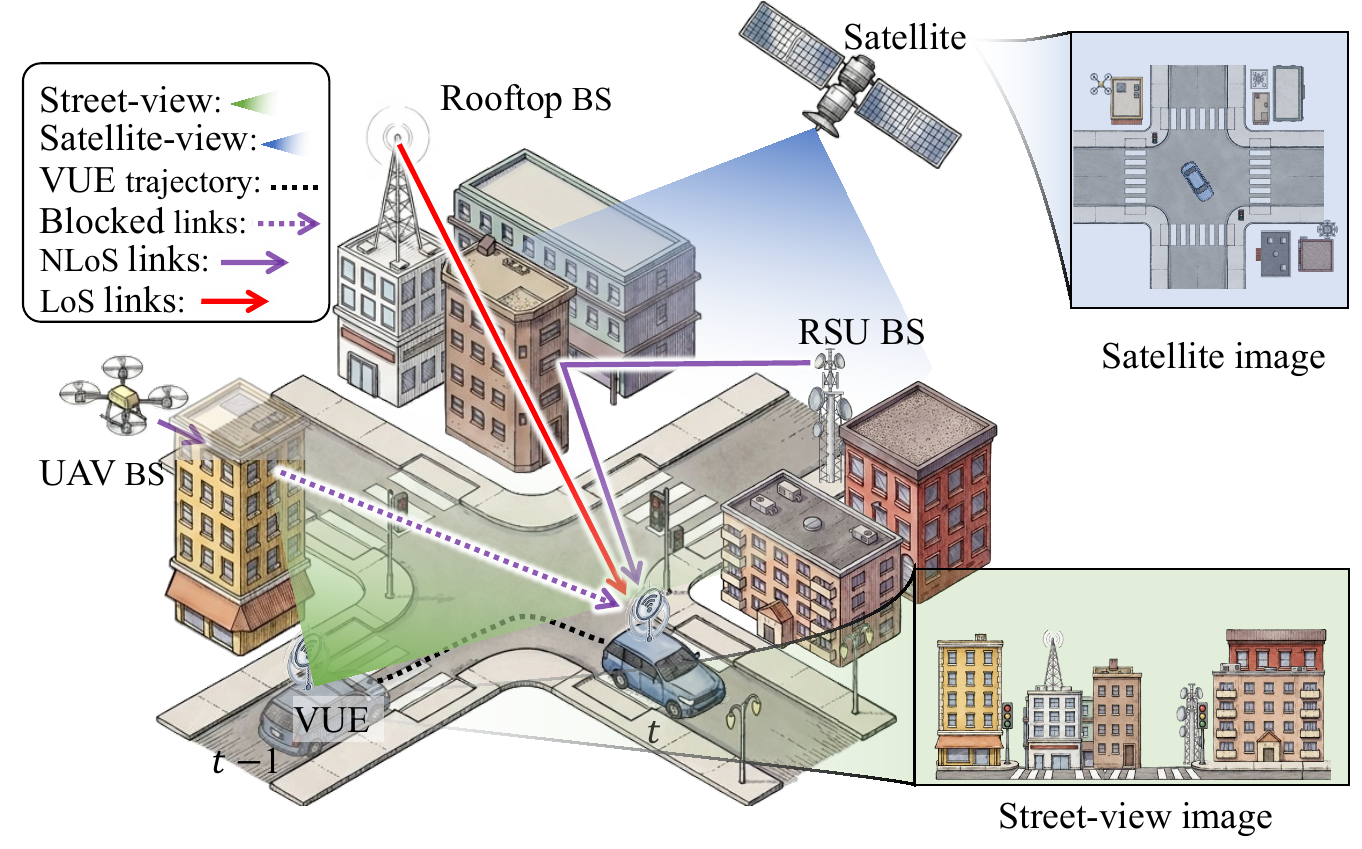}
\vspace{-0.8cm}
\caption{Illustration of the considered urban mmWave V2I scenario.}
\label{scene}
\vspace{-0.5cm}
\end{figure}

We consider a downlink mmWave V2I communication system deployed in an urban road environment, as illustrated in Fig.~\ref{scene}. In this environment, a VUE is equipped with GPS, a camera, and an onboard mmWave transceiver, and is served by one BS from a nearby candidate set $\mathcal{B}\triangleq\{1,\ldots,B\}$, which is spatially distributed across the environment, such as roadside, rooftop, or UAV. The positions of candidate BSs are known to the VUE. At each time slot $t$, the VUE can access satellite images via the on-board navigation system, obtain street-view images via its camera, and observe its locations via GPS record. The VUE interprets the channel knowledge behind these observations and then selects the BS to maximize the transmission rate in the downlink communication at time slot $t+1$. The corresponding decision is sent by the VUE to the target BS during the stable data-transmission stage of slot $t$, before the association is applied in slot $t+1$. The VUE and the target BS then establish the association according to the selected BS index, and perform the subsequent downlink communication using the selected beam-search result.

\subsection{System Model}
\subsubsection{mmWave V2I Channel Model}
Both the BS and the VUE are equipped with uniform planar arrays (UPAs), containing $N\!=\!N_xN_y$ transmit antennas and $M\!=\!M_xM_y$ receive antennas, respectively, with a single radio frequency chain. We note that the UPAs are modeled on $x$-$y$ planes parallel to the ground, with their element indices column-first from the top-left to the bottom-right. The channel between each candidate BS and the VUE is modeled as an orthogonal frequency-division multiplexing (OFDM) mmWave link with carrier frequency $f_c$ and $K$ subcarriers. Following a standard geometric wideband mmWave channel model \cite{rappaportOverviewMillimeterWave2017}, the frequency-domain channel on the $k$-th subcarrier, $k \in \{1, \ldots ,K\}$, denoted by $\mathbf{H}_k\in\mathbb{C}^{M\times N}$, is given by:
\begin{equation}\label{hk}
\mathbf{H}_k =
\sum_{\xi=0}^{\Xi-1}
\sum_{\ell=1}^{L_{\rm p}}
\alpha_\ell
e^{-\boldsymbol{\jmath}\frac{2\pi k}{K}\xi}
d(\xi T_s-\tau_\ell)
\mathbf{r}(\phi_\ell^{\mathrm{A}},\theta_\ell^{\mathrm{A}})
\mathbf{b}^{\mathrm{H}}(\phi_\ell^{\mathrm{D}},\theta_\ell^{\mathrm{D}}).
\end{equation}
where $\Xi$ represents the length of the discrete-time channel impulse response; $L_{\rm p}$ is the number of propagation paths; the exponential term $e^{-\boldsymbol{\jmath}\frac{2\pi k}{K} \xi}$ represents the standard discrete Fourier transform (DFT) operation for OFDM demodulation, and $\boldsymbol{\jmath}=\sqrt{-1}$ is the imaginary unit; $\alpha_\ell$ and $\tau_\ell$ are the complex gain and delay of the $\ell$-th path, respectively; $T_{s}$ is the sampling interval;  $d(\cdot)$ is a raised-cosine pulse to eliminate inter-symbol interference under high-speed transmission; $\mathbf{r}(\phi_l^A,\theta_l^A)\in\mathbb{C}^{M\times1}$ and $\mathbf{b}(\phi_l^D,\theta_l^D)\in\mathbb{C}^{N\times1}$ are the steering vectors of UPA at VUE and BS, respectively. Specifically, for the $\ell$-th path, $\phi_\ell^{\mathrm{A}}$ and $\theta_\ell^{\mathrm{A}}$ are the azimuth and elevation angles of arrival (AOA), respectively. $\phi_\ell^{\mathrm{D}}$ and $\theta_\ell^{\mathrm{D}}$ are the azimuth and elevation angles of departure (AOD), respectively. For a typical $U$-antenna UPA ($U=U_x \times U_y$), the steering vector is given as follows:
\begin{equation}
\mathbf{a}(\phi,\theta)
\triangleq
\mathbf{a}_{y}(\phi,\theta)\otimes\mathbf{a}_{x}(\phi,\theta), \forall \mathbf{a}\!\in\!\{\mathbf{r},\!\mathbf{b}\},
\end{equation}
where 
\begin{align}
\mathbf{a}_{x}(\phi,\theta)
&\triangleq
\left[
1,e^{\boldsymbol{\jmath}\pi\sin\theta\cos\phi},\ldots,
e^{\boldsymbol{\jmath}\pi(U_x-1)\sin\theta\cos\phi}
\right]^{\top},\\
\mathbf{a}_{y}(\phi,\theta)
&\triangleq
\left[
1,e^{\boldsymbol{\jmath}\pi\sin\theta\sin\phi},\ldots,
e^{\boldsymbol{\jmath}\pi(U_y-1)\sin\theta\sin\phi}
\right]^{\top}.
\end{align}

\subsubsection{Transmission Rate under mmWave Channel}
Let $\mathcal{W}^{\mathrm{tx}}\triangleq\{\mathbf{w}_i^{\mathrm{tx}}\}_{i=1}^{C_{\mathrm{tx}}}$ and $\mathcal{W}^{\mathrm{rx}}\triangleq\{\mathbf{w}_j^{\mathrm{rx}}\}_{j=1}^{C_{\mathrm{rx}}}$ denote the transmit and receive analog beam codebooks of the BS and VUE, respectively, with the sizes of $C_{\mathrm{tx}}$ and $C_{\mathrm{rx}}$, where $i$ and $j$ are the corresponding beam indices. The transmit symbol on the $k$-th subcarrier is denoted by $s_k$, with corresponding power $P_k=\mathbb{E}\{|s_k|^2\}$. The received signal at the VUE on the $k$-th subcarrier is given by:
\begin{equation}
y_k=\left(\mathbf{w}_j^{\mathrm{rx}}\right)^{\mathrm{H}}
\mathbf{H}_k
\mathbf{w}_i^{\mathrm{tx}}s_k
+
\left(\mathbf{w}_j^{\mathrm{rx}}\right)^{\mathrm{H}}\mathbf{n}_k,
\label{eq:received_signal_revision}
\end{equation}
where $\mathbf{w}_i^{\mathrm{tx}} \in\mathbb{C}^{N\times1}$ and $\mathbf{w}_j ^{\mathrm{rx}} \in\mathbb{C}^{M\times1}$ are the unit-norm transmit and receive beamforming vectors, respectively. $\mathbf{n}_k\sim\mathcal{CN}(\mathbf{0},\sigma^2\mathbf{I}_{M})$ is the additive white Gaussian noise (AWGN), where $\mathbf{I}_{M}$ is the identity matrix. The transmission rate averaged over $K$ subcarriers is defined as
\begin{equation}\label{eq:rate_revision}
R(i,j)=
\frac{1}{K}\sum_{k=1}^{K}
\log_2\!\left(
1+\frac{P_k}{\sigma^2}
\left|
\left(\mathbf{w}_j^{\mathrm{rx}}\right)^{\mathrm{H}}\mathbf{H}_k\mathbf{w}_i^{\mathrm{tx}}
\right|^2
\right),
\end{equation}
where equal power allocation is applied, i.e., $P_k=P_{\mathrm{tot}}/K$. 

\subsection{Problem Formulation and Practical Challenges}

\begin{figure}[t]
\centering
\includegraphics[width=\linewidth]{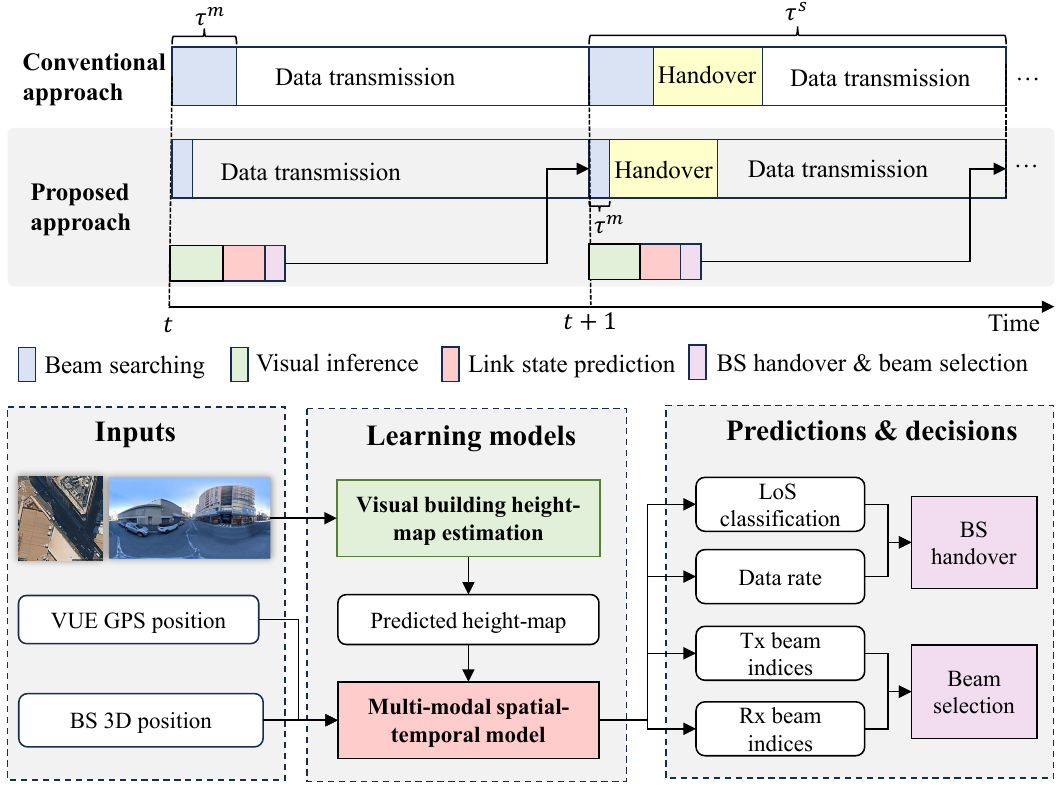}
\vspace{-0.8cm}
\caption{Framework overview. Top: Conventional and proposed BS association procedures; Bottom: Architecture of the proposed framework.}
\label{fig: system overall}
\vspace{-0.5cm}
\end{figure}

\subsubsection{Effective Data-Transmission Rate Maximization}
Typically, each time slot is composed of three time durations for beam searching, handover, and data transmission, respectively, as shown in Fig.~\ref{fig: system overall}. At the beginning of the $t$-th time slot, the VUE first performs the beam searching to obtain the optimal beam index of each BS, which takes $\tau^{\rm m}$. Then, given a BS selection strategy, the VUE decides whether to process the handover based on the result of link quality. The handover duration is denoted by $\tau^{\rm h}$. The remaining time is used to transfer the data between the VUE and the selected BS via the searched beam index. Hence, the data transmission duration at the $t$-th time slot is given by:
\begin{equation}\label{eq:data_duration}
    \tau_t^{\rm d}
    =
    \tau^{\rm s}
    -
    \tau^{\rm m}
    -
    \underbrace{\left(1-\mathbf{o}_{t-1}^{\top}\mathbf{o}_t\right)}_{\delta_t}
    \tau^{\rm h},
\end{equation}
where $\tau^{\rm s}$ is the duration of each time slot, and $\mathbf{o}_t=[o_{1,t},\ldots,o_{B,t}]^{\top}\in\{0,1\}^{B}$ denotes the one-hot association vector at slot $t$, i.e., $o_{b,t}=1$ indicates that the VUE is associated with candidate BS $b$ during the $t$-th time slot and $o_{b,t}=0$ otherwise. We note that the VUE is served by only one BS in each time slot, indicating $\mathbf{1}^{\top}\mathbf{o}_{t}=1$. Therefore, $\delta_t=1-\mathbf{o}_{t-1}^{\top}\mathbf{o}_t$ is the handover indicator, i.e., $\delta_t=1$ represents the selected BS changes from slot $t-1$ to slot $t$, and $\delta_t=0$ otherwise. 

For candidate BS $b$ at slot $t$, let $R_{b,t}(i,j)$ denote the transmission rate in \eqref{eq:rate_revision} evaluated using the channel between BS $b$ and the VUE when the transmit and receive beam indices are $i$ and $j$, respectively. Given the previous association $\mathbf{o}_{t-1}$, the VUE needs to choose the current association and beam-index pair to maximize the effective data transmission in the $t$-th time slot, which can be formulated as
\begin{subequations}\label{eq:slot_problem}
\begin{align}
\left(\mathbf{o}_t^{\star},i_t^{\star},j_t^{\star}\right)
=
\arg\max_{\mathbf{o}_t,i_t,j_t}
&
\frac{\tau_t^{\mathrm{d}}}{\tau^{\mathrm{s}}}
\sum_{b=1}^{B}
o_{b,t}R_{b,t}(i_t,j_t)
\label{eq:slot_objective}
\\
\mathrm{s.t.}\quad
&
\mathbf{o}_t\in\{0,1\}^{B},\quad
\mathbf{1}^{\top}\mathbf{o}_t=1,
\label{eq:slot_assoc_constraint}
\\
&
1\leq i_t\leq C_{\mathrm{tx}},\quad
1\leq j_t\leq C_{\mathrm{rx}}.
\label{eq:slot_beam_constraint}
\end{align}
\end{subequations}
Here, $\tau_t^{\mathrm{d}}/\tau^{\mathrm{s}}$ converts the transmission rate into the effective data-transmission rate after accounting for beam-search and the handover cost induced by switching from $\mathbf{o}_{t-1}$ to $\mathbf{o}_t$. Given the beam index pair $(i_t^\star,j_t^\star)$, the analog beams for the selected BS at slot $t$ are $\mathbf{w}^{\mathrm{tx}}_{i_t^\star}$ and $\mathbf{w}^{\mathrm{rx}}_{j_t^\star}$ from the predefined codebooks.

\subsubsection{Challenges for Solving the Rate Maximization Problem}
Solving \eqref{eq:slot_objective}-\eqref{eq:slot_beam_constraint} by conventional methods is difficult because the association variable $\mathbf{o}_t$ and the beam indices $i_t$ and $j_t$ are coupled. \textbf{Beam-search overhead:} For each candidate BS, evaluating $R_{b,t}(i,j)$ either requires estimating $\mathbf{H}_k$ in \eqref{hk} or sweeping many beam pairs over both codebooks~\cite{tanBeamAlignmentMmWave2024,madhekwanaBeamAlignmentMmWave2025}. The resulting $\tau^{\rm m}$ reduces $\tau_t^{\rm d}$ in \eqref{eq:data_duration}, particularly for large $B$, $C_{\rm tx}$, or $C_{\rm rx}$.
\textbf{Mobility-induced degradation:}
VUE movement relative to blockage rapidly changes the LoS condition, dominant paths, and optimal beams. Channel estimation, therefore, requires substantial pilot overhead, and the resulting estimates or searched beam indices can easily be outdated in these rapidly varying channels~\cite{tanBeamAlignmentMmWave2024,rappaportOverviewMillimeterWave2017}.
\textbf{Handover cost:}
Changing $\mathbf{o}_{t-1}$ to $\mathbf{o}_t$ activates $\delta_t$ and incurs $\tau^{\rm h}$, reducing $\tau_t^{\rm d}$. Thus, selecting the BS with the largest data rate may be suboptimal when its gain cannot offset the handover loss and may also cause frequent switching when multiple BSs have similar rates~\cite {ahnSensingComputerVisionAided2024}. Therefore, beam-search cost, handover cost, and transmission rate must be jointly optimized.

\section{Proposed Vision-Aided Proactive BS Selection and Beam Prediction}\label{sec:LearningFormu}
\subsection{Learning-based Approach Formulation}

To address the challenges for solving \eqref{eq:slot_objective}-\eqref{eq:slot_beam_constraint}, we formulate a proactive learning-based approach, which uses the sensing and mobility information available up to slot $t$ to predict the next-slot link state for each candidate BS before the slot starts. The input includes the VUE trajectory, known candidate-BS locations, and visual environment observations. The output for each BS $b$ includes the predicted LoS indicator $\hat{p}^{\mathrm{LoS}}_{b,t+1}\in\{0,1\}$, the predicted optimal data rate $\hat{R}^{\star}_{b,t+1}$, and the predicted transmit and receive beam rankings $\mathcal{R}^{\mathrm{tx}}_{t+1}\triangleq\{{{i}}\}^{K^{\rm tx}_{\mathrm{beam}}}_{t+1}$, $\mathcal{R}^{\mathrm{rx}}_{t+1}\triangleq\{{{j}}\}^{K^{\rm rx}_{\mathrm{beam}}}_{t+1}$ over $\mathcal{W}^{\mathrm{tx}}$ and $\mathcal{W}^{\mathrm{rx}}$. These learned predictions are used to approximate the same effective-rate objective in \eqref{eq:slot_objective}. Specifically, we first use the predicted optimal data rate $\hat{R}^{\star}_{b,t+1}$ to determine the BS selection at time $t+1$, which can be expressed as 
\begin{subequations}\label{eq:BS_assoc_problem}
\begin{align}
\mathbf{\hat{o}}_{t+1}^{\star}
=
\arg\max_{\mathbf{\hat{o}}_{t+1}}
&
\frac{\tau^{\rm d}_{t+1}}{\tau^{\mathrm{s}}}
\sum_{b=1}^{B}
\hat{o}_{b,t+1}\hat{p}^{\mathrm{LoS}}_{b,t+1}\hat{R}^{\star}_{b,t+1}
\label{eq:slot_objective1}
\\
\mathrm{s.t.}\quad
&
\mathbf{\hat{o}}_{t+1}\in\{0,1\}^{B},\quad
\mathbf{1}^{\top}\mathbf{\hat{o}}_{t+1}=1.
\label{eq:slot_assoc_constraint1}
\end{align}
\end{subequations}
Given $\mathbf{\hat{o}}_{t+1}^{\star}$, we perform the beam search with respect to $\mathcal{R}^{\mathrm{tx}}_{t+1}$ and $\mathcal{R}^{\mathrm{rx}}_{t+1}$ to maximize the effective data transmission in the time slot $t+1$, which is expressed as follows:
\begin{subequations}\label{eq:Beam_search_problem}
\begin{align}
\left(\!{i}_{t+1}^{\star},{j}_{t+1}^{\star}\!\right)
\!\!=\!
\arg\!\!\!\!\max_{{i}_{t+1},{j}_{t+1}}
&
\!\!\!\frac{\tau_{t+1}^{\mathrm{d}}}{\tau^{\mathrm{s}}}\!\!
\sum_{b=1}^{B}\!
{\hat{o}_{b,t+1}^{\star}}\!R_{b,t+1}\!(\!{i}_{t+1},{j}_{t+1}\!)
\label{eq:slot_objective2}
\\
\mathrm{s.t.}\quad
&
i_{t+1}\in\mathcal{R}^{\mathrm{tx}}_{t+1},\quad
j_{t+1}\in\mathcal{R}^{\mathrm{rx}}_{t+1}.
\label{eq:slot_beam_constraint2}
\end{align}
\end{subequations}
Following staged BS/VUE refined beam search procedures~\cite{3gppTR38802}, the beam search duration is
$\tau^{\mathrm{m}}
=(K^{\rm tx}_{\mathrm{beam}}+ K^{\rm rx}_{\mathrm{beam}})\tau_{\mathrm{u}}$,
where $\tau_{\mathrm{u}}$ denotes the unit beam-probing time.

The formulated learning-based approach targets the same transmission rate objective as \eqref{eq:slot_objective}-\eqref{eq:slot_beam_constraint}, while replacing
 unavailable next-slot channel-dependent quantities with predictions and reducing the online search space from exhaustive BS-beam exploration to rate-aware BS selection and ordered beam testing. Using environmental and mobility information available before data transmission in slot $t+1$, the proposed framework proactively determines the BS selection and the beam-search shortlist to maximize the transmission rate in \eqref{eq:slot_problem}.


\subsection{Learning-based Framework}
To address the learning problem of \eqref{eq:slot_objective1}-\eqref{eq:slot_beam_constraint2}, we propose a vision-aided, proactive V2I prediction framework that extracts surrounding environmental information from camera-captured street-view images and a preloaded satellite map from the navigation database for next-slot BS selection and beam prediction. The framework contains two main components: the VBHE and the MMST. 

As illustrated in Fig.~\ref{fig: system overall}, the VBHE first takes a street-view panoramic image and a satellite image as inputs and outputs a building height-map representation that summarizes building footprints and heights around the VUE. The MMST then combines this height map with the historical VUE and candidate-BS coordinates, extracts candidate-specific geometric features along each VUE-BS path, and predicts the next-slot LoS indicator $\hat{p}^{\mathrm{LoS}}_{b,t+1}$, optimal-beam rate $\hat{R}^\star_{b,t+1}$, and the Top-$K^{\rm tx}_{\mathrm{beam}}/K^{\rm rx}_{\mathrm{beam}}$ beam indices, ranked by their predicted beam probabilities, of BS and VUE for each candidate BS $b$, respectively. The architecture and training objective of VBHE and MMST are detailed in Sections~\ref{sec:VBHE} and ~\ref{sec:mmst}, respectively.
\begin{figure*}[t]  
\centering  
\includegraphics[width=\linewidth]{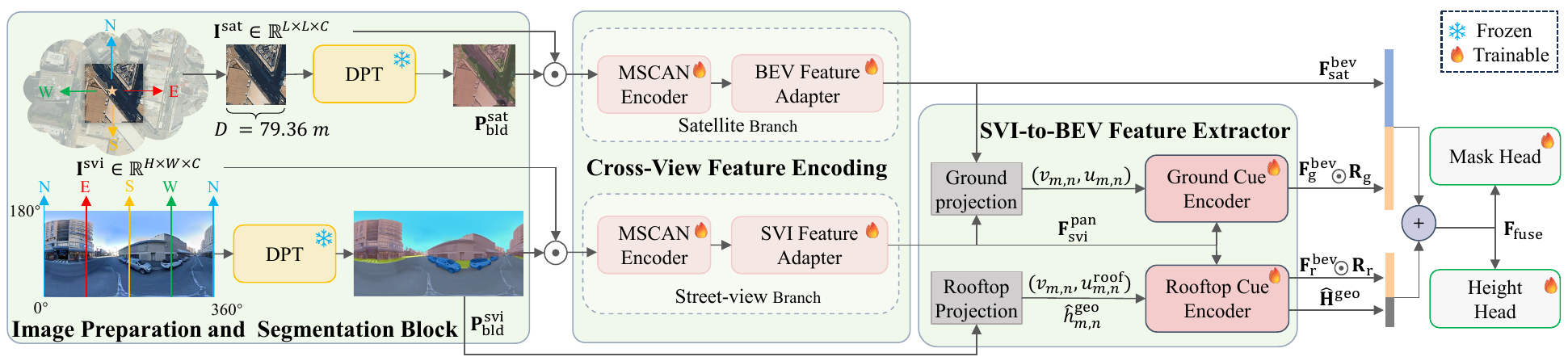}
\vspace{-0.8cm}
\caption{Overall architecture of the proposed cross-view VBHE model.}
\label{cv-model}
\vspace{-0.5cm}
\end{figure*}

\section{Visual Building Height-Map Estimation Model}~\label{sec:VBHE}
The VBHE estimates the building footprints and height map around the VUE. The height map encodes building geometry that is informative for blockage, LoS/NLoS conditions, and possible reflected paths. This geometric information is used by the subsequent MMST for BS selection and beam-index prediction.
As shown in Fig.~\ref{cv-model}, VBHE consists of four components: 1)~an image preparation and segmentation block; 2)~a cross-view feature encoding block; 3)~SVI-to-BEV feature extractor; and 4)~a cross-view fusion and output block.

\subsection{Image Preparation and Segmentation Block}
\subsubsection{Image Preparation}
At each time slot, the VUE provides the satellite image, denoted by $\mathbf{I}^{\mathrm{sat}}_t\!\in\!\mathbb{R}^{L \times L \times C}$, and the street-view panorama image, denoted by $\mathbf{I}^{\mathrm{svi}}_t\!\in \!\!\mathbb{R}^{H\times W \times C}$, as inputs to the VBHE. Specifically, the satellite image \(\mathbf{I}^{\mathrm{sat}}_t\) is a North-aligned RGB patch centered at the VUE position \(\mathbf{p}^{\rm UE}_t\), covering a \(D\!\times\!\!D~\mathrm{m}^2\) area. The street-view panorama image $\mathbf{I}^{\mathrm{svi}}_t$ is obtained using an onboard $360^\circ$ camera mounted $h_{\rm c}$ m above the ground. This image is presented in the equirectangular format, where the horizontal axis uniformly covers the full $360^\circ$ azimuth range and the vertical axis uniformly covers the $180^\circ$ elevation range. The left boundary of $\mathbf{I}_t^{\mathrm{svi}}$ corresponds to true North to align the street-view panorama with $\mathbf{I}^{\mathrm{sat}}_t$. For ease of notation, we omit the subscript $t$ in the subsequent analysis.

\subsubsection{Image Segmentation}
Since the VBHE aims to estimate building footprints and heights, masking non-building regions, such as roads, sky, vegetation, and vehicles, reduces background interference and helps the subsequent feature extractor focus on building-related visual cues. Hence, a DPT-hybrid building segmentation module~\cite{ranftlVisionTransformersDense2021} is deployed to identify building regions from both $\mathbf{I}^{\mathrm{sat}}$ and $\mathbf{I}^{\mathrm{svi}}$.\footnote{DPT modules \cite{ranftlVisionTransformersDense2021} are trained offline on the semantic segmentation dataset. The parameters are frozen during training and deployment, so the block introduces no additional learnable parameters into the proposed VBHE.} 
Specifically, we build the building probability map $\mathbf{P}_{\mathrm{bld}}^{\nu}$ by 
\begin{equation}
    \mathbf{P}_{\mathrm{bld}}^{\nu}
=
\operatorname{softmax}_{c_{\mathrm{bld}}}\!\left(f^{\mathrm{DPT}}(\mathbf{I}^{\nu};\boldsymbol{\theta}^{\mathrm{D}})\right)
,\nu\in\{\mathrm{sat},\mathrm{svi}\},
\end{equation}
where $\mathbf{P}_{\mathrm{bld}}^{\mathrm{sat}}\in[0,1]^{L\times L}$, $\mathbf{P}_{\mathrm{bld}}^{\mathrm{svi}}\in[0,1]^{H\times W}$, and $f^{\mathrm{DPT}}(\cdot;\boldsymbol{\theta}^{\mathrm{D}})$ denotes the DPT-Hybrid model with parameters $\boldsymbol{\theta}^{\mathrm{D}}$. The softmax operation is applied over the semantic-class dimension of the DPT output, and $c_{\mathrm{bld}}$ denotes the building-class index. 
Then, the building-masked image is $\tilde{\mathbf{I}}^{\nu}\!=\!\mathbf{P}_{\mathrm{bld}}^\nu\odot\mathbf{I}^{\nu}$ to suppress non-building regions for the subsequent feature encoding.

\begin{figure*}[t]  
\centering  
\includegraphics[width=\linewidth]{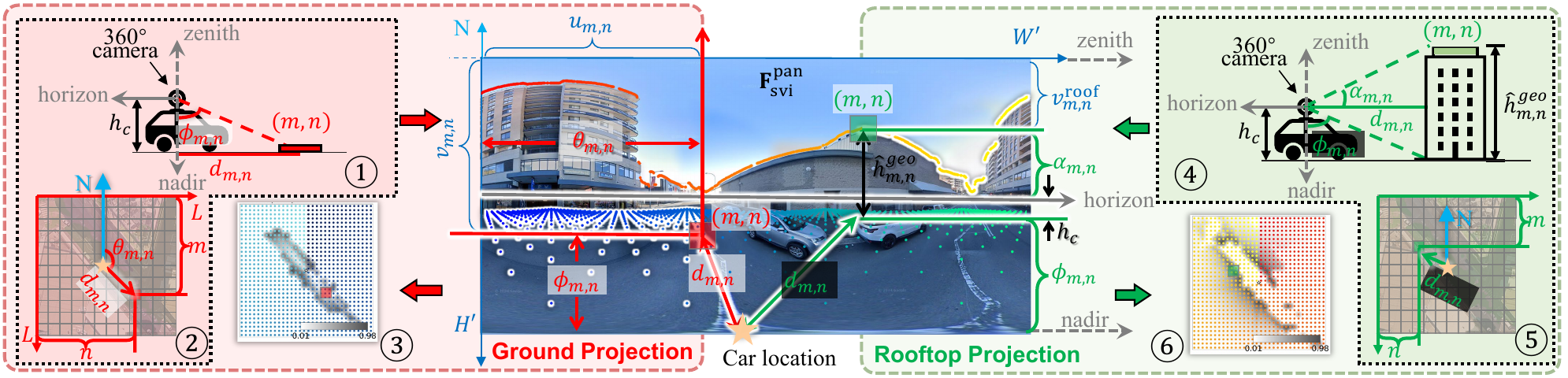}
\vspace{-0.8cm}
\caption{Illustration of the ground and rooftop projection. For ground projection: \textcircled{\raisebox{-0.9pt}{1}} the vertical viewing angle $\phi_{m,n}$ and horizontal distance $d_{m,n}$ of BEV cell $(m,n)$ relative to the camera; \textcircled{\raisebox{-0.9pt}{2}} the azimuth angle $\theta_{m,n}$ and distance $d_{m,n}$ of $(m,n)$ on BEV domain; and \textcircled{\raisebox{-0.9pt}{3}} the sampled location from the panorama to the $\tilde{\mathbf{F}}_{\rm g}^{\rm bev}$ with corresponding  $\mathbf{R}_{\rm g}$. For rooftop projection: \textcircled{\raisebox{-0.9pt}{4}} the rooftop elevation angle $\alpha_{m,n}$ of BEV cell $(m,n)$ relative to the camera; \textcircled{\raisebox{-0.9pt}{5}} the azimuth angle $\theta_{m,n}$ and distance $d_{m,n}$ of $(m,n)$ on BEV domain; and \textcircled{\raisebox{-0.9pt}{6}} sampled location from the panorama to the $\tilde{\mathbf{F}}_{\rm r}^{\rm bev}$ with corresponding reliability $\mathbf{R}_{\rm r}$.}
\label{bhe}
\vspace{-0.5cm}
\end{figure*}

\subsection{Cross-View Feature Encoding Block}
The cross-view feature encoding block adopts a two-branch architecture~\cite{yeSGBEVSatelliteGuidedBEV2024}. As shown in Fig.~\ref{cv-model}, two parallel MSCAN encoders extract multi-scale pyramid features from $\tilde{\mathbf{I}}^{\nu}$~\cite{guo2022segnext}. The extracted pyramid features are then fed into subsequent view-specific feature adapters to map them into a compact branch-level representation.

\subsubsection{MSCAN Encoder}
MSCAN encoders use the multi-scale convolutional attention design to extract multi-scale pyramid features. These features capture local details, such as building boundaries and global contextual cues, which are important for cross-view feature representation.
The multi-scale pyramid features $\{\mathbf{c}^{\nu}_{l}\}_{l=1}^{4}$ with the corresponding downsampling ratios $\{1/4,1/8,1/16,1/32\}$ are extracted by
\begin{equation}
\{\mathbf{c}^{\nu}_{l}\}_{l=1}^{4}
=
f^{\mathrm{E}}
\bigl(\tilde{\mathbf{I}}^{\nu};\boldsymbol{\theta}^{\mathrm{E},\nu}\bigr),
\quad
\nu\in\{\mathrm{sat},\mathrm{svi}\},
\label{eq:mscan_encoder}
\end{equation}
where $f^{\mathrm{E}}(\cdot;\boldsymbol{\theta}^{\mathrm{E},\nu})$ denotes the MSCAN encoder for view $\nu$, with trainable parameters $\boldsymbol{\theta}^{\mathrm{E},\nu}$.

\subsubsection{Cross-view Adapter}
The two view-specific adapters follow a similar lightweight multi-scale fusion structure. For each adapter, an MLP is first applied to project the feature into the channel dimension $D_b$. The projected features are then resized to a branch-specific reference resolution and fused through convolutional layers. 
Finally, the BEV adapter outputs $\mathbf{F}_{\mathrm{sat}}^{\mathrm{bev}}\in\mathbb{R}^{D_b\times L\times L}$, which is directly treated as the representation in the BEV domain, whereas the SVI adapter outputs $\mathbf{F}_{\mathrm{svi}}^{\mathrm{pan}}\in\mathbb{R}^{D_b\times H'\times W'}$ in the panorama domain.
The two adapters are
\begin{equation}
\mathbf{F}^{\mathrm{bev}}_{\mathrm{sat}}
=
f^{\mathrm{A}}
\bigl(\{\mathbf{c}^{\mathrm{sat}}_{l}\}_{l=1}^{4};\boldsymbol{\theta}^{\mathrm{A},\mathrm{sat}}\bigr),
\label{eq:sat_adapter}
\end{equation}
\begin{equation}
\mathbf{F}^{\mathrm{pan}}_{\mathrm{svi}}
=
f^{\mathrm{A}}
\bigl(\{\mathbf{c}^{\mathrm{svi}}_{l}\}_{l=2}^{4};\boldsymbol{\theta}^{\mathrm{A},\mathrm{svi}}\bigr),
\label{eq:svi_adapter}
\end{equation}

For the satellite branch, all pyramid levels are used because the satellite features are already in the top-down BEV domain and provide complementary layout cues at different scales. In contrast, the street-view features need to be projected into the BEV domain, and thus should preserve sufficient spatial detail while maintaining a receptive field large enough to provide reliable geometric and semantic cues. The finest level $l=1$ mainly captures highly local panorama textures, which are sensitive to perspective distortion and provide limited reliable information for BEV projection, while also increasing the panorama-to-BEV sampling cost. Therefore, using levels $l=2,3,4$ for the street-view adapter provides a better balance between spatial resolution, contextual representation, and computational efficiency for subsequent feature extractors.

\subsection{SVI-to-BEV Feature Extractor}
Since both the satellite representation and the height map are defined in the BEV domain, the street-view representation $\mathbf{F}^{\mathrm{pan}}_{\mathrm{svi}}$ in~\eqref{eq:svi_adapter}, which remains in the panorama domain, must be projected into the satellite BEV coordinate system before cross-view fusion. We therefore explore the mapping between the panorama and BEV coordinate systems to align the panorama-domain features $\mathbf{F}^{\mathrm{pan}}_{\mathrm{svi}}$ with BEV features $\mathbf{F}^{\mathrm{bev}}_{\mathrm{sat}}$. Without this mapping, features at the same tensor index in representation $\mathbf{F}^{\mathrm{pan}}_{\mathrm{svi}}$ and $\mathbf{F}^{\mathrm{bev}}_{\mathrm{sat}}$ would correspond to different physical locations, leading to spatially inconsistent fusion.

As illustrated in Fig.~\ref{cv-model}, the ground and rooftop projections first find the relationship between each cell in the BEV-domain $\mathbf{F}^{\mathrm{bev}}_{\mathrm{sat}}$ with the corresponding ground coordinate and rooftop coordinate on the panorama-domain street-view feature map $\mathbf{F}^{\mathrm{pan}}_{\mathrm{svi}}$, respectively. 
The mappings are then fed to ground cue encoder and rooftop cue encoder, which resample the ${\mathbf{F}}^{\mathrm{pan}}_{\mathrm{svi}}$ to generate the BEV-aligned street-view ground feature $\tilde{\mathbf{F}}^{\mathrm{bev}}_{\mathrm{g}}\in\mathbb{R}^{D_b\times L\times L}$ and rooftop feature $\tilde{\mathbf{F}}^{\mathrm{bev}}_{\mathrm{r}}\in\mathbb{R}^{D_b\times L\times L}$, respectively. 

\subsubsection{Ground and Rooftop Projection}
As shown in Fig.~\ref{bhe}, the ground and rooftop projections are based on the geometric relationship between the camera and the pixel on the ground plane or on the rooftop, respectively. 

\underline{Ground Projection:}
For each spatial BEV cell $(m,n)$ on the ground plane, where $m,n\in\{1,\ldots,L\}$, its location corresponds to the row and column coordinates on panorama-domain image $(v_{m,n}, u_{m,n})$, where $v_{m,n}\in[0,H']$ and $u_{m,n}\in[0,W']$ are defined by the standard equirectangular projection (ERP) properties~\cite{coorsSphereNetLearningSpherical2018}, i.e., $v=0$ corresponds to the zenith, $v=H'/2$ corresponds to the horizon, and $v=H'$ corresponds to the nadir. Therefore, the coordinates are computed as
\begingroup
\begin{equation}
    u_{m,n} = \frac{\theta_{m,n}}{2\pi}\, W',\qquad
     v_{m,n} = H'-\frac{\phi_{m,n}}{\pi} H',
    \label{eq:uv}
\end{equation}
\endgroup
where $\theta_{m,n}$ denotes the azimuth angle from the VUE position to the BEV cell $(m,n)$, measured with respect to the north direction; $\phi_{m,n}$ is the vertical viewing angle from the camera center to the building base point on the ground plane. As shown in \circlednum{2} in Fig.~\ref{bhe}, the angle $\theta_{m,n}\in[0,2\pi)$ is computed as follows.
\begingroup
\begin{equation}
    \theta_{m,n}
    \!=\!\operatorname{mod}\big[
    \mathrm{atan2}
    \Bigl(
    \underbrace{(n-\frac{L}{2})\frac{D}{L}}_{\Delta y_n},
    \underbrace{-(m-\frac{L}{2})\frac{D}{L}}_{-\Delta x_m}
    \Bigr),
2\pi \big].
    \label{eq:theta}
\end{equation}
\endgroup
where $\Delta x_m$ and $\Delta y_n$ are the metric ground-plane offsets from cell $(m,n)$ to the BEV-grid center, which corresponds to the current VUE/camera location. The factor $D/L$ is the metric scale factor, with units of m/pixel, that converts grid offsets to distances in meters. The vertical viewing angle $\phi_{m,n}$, as shown in \circlednum{1} in Fig.~\ref{bhe}, is computed as
\begingroup
\begin{equation}
    \phi_{m,n} = \arctan\!\left({d_{m,n}}/{h_{\rm c}}\right),
    \label{eq:phi}
\end{equation}
\endgroup
where $h_{\rm c}$ is the camera height, and $d_{m,n}$ is the horizontal ground distance from the camera location to the BEV cell in meters, computed by $d_{m,n} = \sqrt{{\Delta x_m}^2 + {\Delta y_n}^2}$.

\underline{Rooftop Projection:}
For each spatial BEV cell $(m,n)$ on the rooftop, its location corresponds to the panorama feature coordinates $(v^{\mathrm{roof}}_{m,n},u_{m,n})$. The horizontal image coordinate $u_{m,n}$ is the same as in~\eqref{eq:uv}, while the vertical coordinate $v^{\mathrm{roof}}_{m,n} \in [0, H']$ is determined by finding the topmost row in $\mathbf{P}_{\mathrm{bld}}^{\mathrm{svi}}$, whose building probability exceeds 0.5 at fixed column index $\lfloor u_{m,n} \rfloor \frac{W}{W'} $. Specifically,
\begin{equation}
v^{\mathrm{roof}}_{m,n}
=
\begin{cases}
\min\mathcal{F}_{m,n}, & \mathcal{F}_{m,n}\neq\emptyset,\\
H'/2, & \mathcal{F}_{m,n}=\emptyset,
\end{cases}
\label{eq:rooftop_projection_row}
\end{equation}
where
\begin{small}
\begin{equation}
\mathcal{F}_{m,n}\!
    \!\triangleq\bigl\{v \!\in\! \{0,\ldots,\frac{H'}{2}\} \!\!\mid \!\mathbf{P}_{\mathrm{bld}}^{\mathrm{svi}}[ v\frac{H}{H'},\lfloor u_{m,n} \rfloor \frac{W}{W'}] >0.5\bigr\}.
\end{equation}
\end{small}
This definition explicitly searches over the vertical coordinate $v$ while keeping the horizontal panorama column $\lfloor u_{m,n} \rfloor \frac{W}{W'} $ fixed by the BEV cell direction.

We can then compute the rooftop height prior geometrically. As shown in \circlednum{4} in Fig.~\ref{bhe}, the prior building rooftop height at BEV cell $(m,n)$, denoted by $\hat{h}^{\mathrm{geo}}_{m,n}$, is estimated as:
\begin{equation}
    \hat{h}^{\mathrm{geo}}_{m,n} = h_{\rm c} + d_{m,n} \tan(\alpha_{m,n}),
    \label{eq:height_geo}
\end{equation}
where $\alpha_{m,n}$ is the elevation angle from the camera to the rooftop along the azimuth direction $\theta_{m,n}$, which is computed by mapping the rooftop row to its angle in the equirectangular projection, given by $\alpha_{m,n} =  \pi\left({H'^{}/2 - v^{\mathrm{ roof}}_{m,n}}\right)/{H'}^{}$.

\subsubsection{Ground Cue Encoder}
After obtaining all the sampling coordinates {$(v_{m,n},u_{m,n})$} for each BEV cell $(m,n)$ on the ground plane, the BEV-aligned street-view ground feature $\tilde{\mathbf{F}}^{\mathrm{bev}}_{\mathrm{g}}$ is obtained by sampling $\mathbf{F}^{\mathrm{pan}}_{\mathrm{svi}}$ at these continuous coordinates, i.e.,  $\forall m,n\in\{1,\ldots L\}$,
\begin{equation}
    \tilde{\mathbf{F}}^{\mathrm{bev}}_{\mathrm{g}}[:,m,n]
=\mathbf{F}^{\mathrm{pan}}_{\mathrm{svi}}[:,\lfloor v_{m,n}\rfloor,\lfloor u_{m,n}\rfloor].
\end{equation}
The resampled ground feature $\tilde{\mathbf{F}}^{\mathrm{bev}}_{\mathrm{g}}$ is then fed into an adapter $f^{\mathrm{G}}(\cdot;\boldsymbol{\theta}^{\mathrm{G}})$ to extract ground-plane-aligned street-view cues and refine the projected representation:
$
    \mathbf{F}^{\mathrm{bev}}_{\mathrm{g}}
    =
    f^{\mathrm{G}}\!\left(
    \tilde{\mathbf{F}}^{\mathrm{bev}}_{\mathrm{g}};
    \boldsymbol{\theta}^{\mathrm{G}}
    \right).
$
 The resulting BEV-aligned ground-plane feature $\mathbf{F}^{\mathrm{bev}}_{\mathrm{g}}\in \mathbb{R}^{D_b\times L\times L}$ is then concatenated with the satellite BEV feature $\mathbf{F}^{\mathrm{bev}}_{\mathrm{sat}}$ for subsequent cross-view fusion.

In addition, we construct a ground reliability matrix $\mathbf{R}_{\rm g}\in[0,1]^{L\times L}$ from the ground regions identified in the building probability map $\mathbf{P}_{\mathrm{bld}}^{\mathrm{sat}}$. As shown in \circlednum{3} in  Fig.~\ref{bhe}, $\mathbf{R}_{\rm g}$ assigns higher reliability to ground cells and lower reliability to building cells, since the ground-plane projection is valid only for ground regions. The reliability is highest at building–ground boundaries, encouraging the subsequent neural network to focus on building footprint information.

\subsubsection{Rooftop Cue Encoder}
Similar to the ground cue encoder, the BEV-aligned street-view rooftop feature $\tilde{\mathbf{F}}^{\mathrm{bev}}_{\mathrm{r}}$ is obtained by sampling $\mathbf{F}^{\mathrm{pan}}_{\mathrm{svi}}$ at the rooftop coordinates {$(v^{\rm roof}_{m,n}, u_{m,n})$}, i.e., $\forall m,n\in\{1,\ldots L\}$,
\begin{equation}
   \tilde{\mathbf{F}}^{\mathrm{bev}}_{\rm r}[:,m,n]
   =
   \mathbf{F}^{\mathrm{pan}}_{\mathrm{svi}}
   \left[:, \left\lfloor v^{\mathrm{roof}}_{m,n}\right\rfloor,
   \left\lfloor u_{m,n}\right\rfloor\right].
   \label{eq:roof_app}
\end{equation}
The sampled rooftop feature $\tilde{\mathbf{F}}^{\mathrm{bev}}_{\mathrm{r}}$ is then fed into a learnable rooftop adapter $f^{\mathrm{R}}(\cdot;\boldsymbol{\theta}^{\mathrm{R}})$ to extract compact rooftop-aware features:
\begin{equation}
    \mathbf{F}^{\mathrm{bev}}_{\mathrm{r}}
    =
    f^{\mathrm{R}}\!\left(
    \tilde{\mathbf{F}}^{\mathrm{bev}}_{\mathrm{r}};
    \boldsymbol{\theta}^{\mathrm{R}}
    \right)
    \in \mathbb{R}^{D_r\times L\times L}.
    \label{eq:roof_cue_encoder}
\end{equation}

We also construct a rooftop-correspondence reliability matrix, denoted by $\mathbf{R}_{\rm r}\!\in\![0,1]^{L\times L}$, from the building probability $\mathbf{P}_{\mathrm{bld}}^{\mathrm{sat}}$. As shown in \circlednum{6} in Fig.~\ref{bhe}, $\mathbf{R}_{\rm r}$ assigns higher reliability to the nearby building cells and lower reliability to ground cells, since the rooftop projection is valid only for building regions. The reliability is highest at building--ground boundaries of the BEV domain, encouraging the subsequent neural network to focus on building-top information. 
In order to further improve the rooftop feature, we also incorporate a geometric rooftop height prior matrix $\hat{\mathbf{H}}^{\rm geo}$ given by $\hat{\mathbf{H}}^{\mathrm{geo}}[m,n]\!=\!\hat{h}^{\mathrm{geo}}_{m,n}$.
The rooftop-aware feature $\mathbf{F}^{\mathrm{bev}}_{\mathrm{r}}$, the geometric height prior $\hat{\mathbf{H}}^{\mathrm{geo}}$, and the reliability matrix $\mathbf{R}_{\rm r}$ are the height-related cues for the subsequent cross-view fusion.

\subsection{Cross-view Fusion and Output Block}
The final cross-view fusion feature is obtained by concatenating the satellite and street-view BEV features and then adding the rooftop-aware feature through 
\begin{equation}
\mathbf{F}_{\mathrm{fuse}} \!\!=\! [\mathbf{F}^{\mathrm{bev}}_{\mathrm{sat}};\; (\mathbf{F}^{\mathrm{bev}}_{\mathrm{g}} \odot   \mathbf{R}_{\rm g})]\! + \! f^{\rm UP}\!([(\mathbf{F}^{\mathrm{bev}}_{\mathrm{r}}\odot \mathbf{R}_{\rm r});\; \hat{\mathbf{H}}^{\mathrm{geo}} ]; {\bm \theta}^{\rm up}),
    \label{eq:fuse}
\end{equation}
where $[\cdot;\cdot]$ denotes channel-wise concatenation, $\mathbf{F}_{\mathrm{fuse}}\in\mathbb{R}^{2D_b\times L\times L}$ contains both satellite-view and street-view information and building height prior in the same BEV coordinates. $f^{\rm UP}(\cdot;\boldsymbol{\theta}^{\mathrm{up}})$ is the convolutional layer that maps the concatenated rooftop-aware feature from $(D_r+1)\times L\times L$ to the same channel dimension $2D_b\times L\times L$.

As shown in Fig.~\ref{cv-model}, the fused feature $\mathbf{F}_{\mathrm{fuse}}$ is fed into two task-specific heads for footprint segmentation and height regression, respectively, and is expressed as
\begin{align}\label{eq:mask_height_head}
    \hat{\mathbf{M}}
    &=\mathbb{I} \big(\underbrace{\operatorname{softmax}
   ( f^{\mathrm{M}}
    (\mathbf{F}_{\mathrm{fuse}};\boldsymbol{\theta}^{\mathrm{M}})
    )}_{\tilde{\mathbf{P}}^{\mathrm{M}}}>0.5\big),
    \\
    \hat{\mathbf{H}}\! &=\!\operatorname{softplus}\left(f^{\mathrm{B}}\!\left(\mathbf{F}_{\mathrm{fuse}};\boldsymbol{\theta}^{\mathrm{B}}\right)\right)\!.
\end{align}
{Here, $\tilde{\mathbf{P}}^{\mathrm{M}}$ denotes the predicted building footprint probability map and is used for footprint supervision. $\hat{\mathbf{M}}$ is the predicted binary footprint mask, where $\hat{\mathbf{M}}[m,n]=0$ indicates ground/background and $\hat{\mathbf{M}}[m,n]=1$ indicates building at BEV cell $(m,n)$. }$\hat{\mathbf{H}}$ is the height map that contains the non-negative physical height values of each cell $(m,n)$.

\subsection{Training Objective of VBHE}
To train the proposed VBHE, we use the ground-truth building height and footprint mask as the labels. The loss function is defined as the weighted sum of mask segmentation loss $\mathcal{L}_m$, height regression loss $\mathcal{L}_h$:
\begin{equation}\label{eq:total_loss}
\mathcal{L}\bigl({\bm \theta}^{\mathrm{VBHE}}\bigr)
= \lambda_m\mathcal{L}_m+\lambda_h\mathcal{L}_h,
\end{equation}
where ${\bm \theta}^{\mathrm{VBHE}} \triangleq \{{\bm \theta}^{\mathrm{B}},{\bm \theta}^{\mathrm{M}},{\bm \theta}^{\mathrm{G}},{\bm \theta}^{\mathrm{R}},{\bm \theta}^{\mathrm{up}},{\bm \theta}^{\mathrm{A,svi}},{\bm \theta}^{\mathrm{A,sat}},{\bm \theta}^{\mathrm{E,sat}},\\{\bm \theta}^{\mathrm{E,svi}}\}$, and $\lambda_m,\lambda_h$ are the relative weights. 
The mask segmentation loss $\mathcal{L}_m$ supervises footprint prediction by combining cross-entropy, Dice~\cite{milletari2016vnet}, and GIoU~\cite{rezatofighi2019giou} losses between the predicted footprint probability map $\tilde{\mathbf{P}}^{\mathrm{M}}$ and the ground-truth mask $\mathbf{M}$. These complementary terms improve pixel-wise classification, foreground-background balance, and regional shape agreement. The height regression loss $\mathcal{L}_h$ focuses on precise height estimation, and is computed as the smooth-$\ell_1$ loss~\cite{girshick2015fastrcnn} between the predicted $\hat{\mathbf{H}}$ and the ground-truth height map $\mathbf{H}$.

\section{Proposed {Multi-Modal Multi-task Spatial-Temporal} Model}\label{sec:mmst}
The MMST model estimates the next-slot link state for each candidate BS by leveraging the geometry information from the VUE trajectory and the VBHE-reconstructed building height map. At the $t$-th time slot, the input of MMST includes the predicted building height map $\hat{\mathbf{H}}_t$, VUE location $\mathbf{p}^{\mathrm{UE}}\!=\!\!(L/2,L/2)$ on the map, and the $b$-th BS location $\mathbf{p}^{\mathrm{BS}}_{b,t}=\!(m^{\mathrm{BS}}_{b,t},n^{\mathrm{BS}}_{b,t}); \forall b \in \mathcal{B}$ on the map. The prediction of MMST includes the next-slot LoS indicator $p^{\mathrm{LoS}}_{b,t+1}$, the predicted optimal data rate $\hat{R}^\star_{b,t+1}$, and the predicted beam rankings $\mathcal{R}^{\mathrm{tx}}_{t+1}$, $\mathcal{R}^{\mathrm{rx}}_{t+1}$ of the $b$-th candidate BS, defined in Section~\ref{sec:LearningFormu}. 
As shown in Fig.~\ref{mmst_arch}, MMST consists of three blocks: (1) a spatial feature encoder, (2) a temporal feature encoder, and (3) a prediction block and training objective, which are elaborated as follows.

\subsection{Spatial Encoder}
The spatial encoder extracts two complementary types of features: (i) the height-map context feature that contains the surrounding building geometry information; (ii) the LoS feature that characterizes the propagation condition along each UE--BS path.

\subsubsection{Height-Map Context Feature Extraction}
The predicted height map contains rich link-related environmental information, indicating building-induced blockage, LoS and NLoS propagation conditions, and surrounding structures that may support reflected paths. To extract these propagation-relevant features from $\hat{\mathbf{H}}_t$, we employ a lightweight convolutional encoder at time slot $t$ 
\begin{equation}
\mathbf{c}_t
=
f^{\mathrm{CNN}}\!\left(
\hat{\mathbf{H}}_t;
\boldsymbol{\theta}^{\mathrm{CNN}}
\right)
\in\mathbb{R}^{d_m},
\label{eq:height_map_encoder}
\end{equation}
where $\mathbf{c}_t$ is the height-map context embedding with dimension $d_m$. The encoder $f^{\mathrm{CNN}}$ comprises four cascaded convolutional blocks. Each block consists of a $3\!\times\!3$ convolutional layer, followed by batch normalization, a ReLU activation, and $\!2\times\!2$ max pooling. Finally, a compact geometric feature is obtained by applying global average pooling and a linear projection.

\subsubsection{LoS Feature Extraction}
\begin{figure}[t]
\centering
\includegraphics[width=\linewidth]{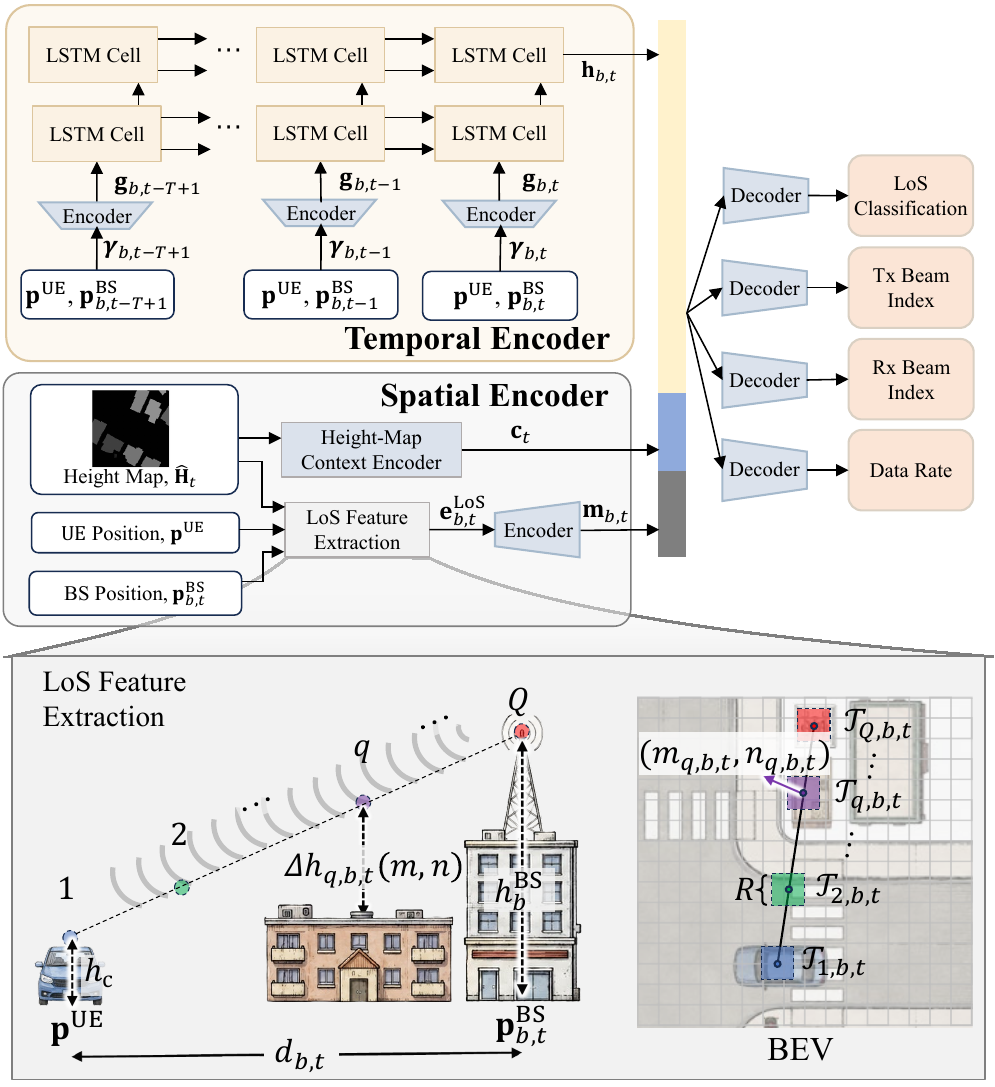}
\vspace{-0.8cm}
\caption{Architecture of the proposed MMST model.}
\label{mmst_arch}
\vspace{-0.5cm}
\end{figure}

The LoS features capture the propagation condition along each VUE-BS path, which is a critical factor for LoS classification. 
As shown in Fig.~\ref{mmst_arch}, we sample $Q$ points along the line segment between the VUE and the BS, where the $q$-th sampled point at $t$-th time slot is indexed by $(m_{q,b,t},n_{q,b,t})$ in height map $\hat{\mathbf{H}}_t$, with $m_{q,b,t},n_{q,b,t}\in\{1,\ldots,L\}$. The height of the $q$-th sampled point towards BS is 
\begin{equation}
h_{q,b}=
h_{\rm c}
+
\frac{q-1}{Q-1}
\left(
h^{\mathrm{BS}}_{b}
-
h_{\rm c}
\right),
\quad
q\in\{1,\ldots,Q\}.
\label{eq:ray_height}
\end{equation}
At each sampled point $(m_{q,b,t},n_{q,b,t})$, we define a local square neighborhood centered at $(m_{q,b,t},n_{q,b,t})$ with side length $R$ pixels as
$\mathcal{T}_{q,b,t}\!=\!\!\{(m,n)\! \mid\!
|m\!-\!m_{q,b,t}|\!\leq\! {(R\!-\!1)/2},
|n\!-\!n_{q,b,t}|\!\leq\! {(R\!-\!1)/2}
\}$.

For each $(m,n)\in\mathcal{T}_{q,b,t}$, the LoS height margin is defined as the height difference between the sampled point and the $(m,n)$-th pixel in the height map, given by $
\Delta h_{q,b,t}(m,n)
\triangleq h_{q,b}
-   \hat{\mathbf{H}}_t[m,n].
$ After computing the height margins for all pixels, we have $\mathcal{H}\triangleq\{\Delta h_{q,b,t}(m,n)\}_{(m,n)\in\mathcal{T}_{q,b,t}}$. 

Then, the LoS feature vector for the
\(q\)-th sampled point towards the \(b\)-th BS at time slot \(t\) is given by
\begin{equation}
\mathbf{e}_{q,b,t}\!
\triangleq\!\!
\left[
\operatorname{mean}(\mathcal H),
\min(\mathcal H),
\left\|\mathbf{1}_{\{\mathcal H<0\}}\right\|_0/|\mathcal{H}|,
\operatorname{std}(\mathcal H)
\right]^{\!\top}\!\!,
\label{eq:local_stat_operator}
\end{equation}
which includes the average height margin, minimum height margin, 
blockage ratio, and height-margin variation, respectively. 

The LoS features $\mathbf{e}_{q,b,t}$ are calculated for all $Q$ sampled points and are concatenated to form the final LoS feature vector for the $b$-th BS at $t$-th time slot: 
$\mathbf{e}_{b,t}^{\mathrm{LoS}}
=
[\mathbf{e}_{1,b,t}^{\top},\ldots,\mathbf{e}_{Q,b,t}^{\top}, 
]^{\top}
\in
\mathbb{R}^{4Q}.
$ Then $\mathbf{e}_{b,t}^{\mathrm{LoS}}$ is transformed to $\mathbf{m}_{b,t}\in\mathbb{R}^{d_h}$ by an MLP encoder, with corresponding parameters $\boldsymbol{\theta}^{\mathrm{sp}}$, and is expressed as $\mathbf{m}_{b,t}=f^{\mathrm{sp}}\!(\mathbf{e}_{b,t}^{\mathrm{LoS}};\boldsymbol{\theta}^{\mathrm{sp}})$.

\subsection{Temporal Encoder}
The temporal encoder includes a relative-position feature extractor and a two-layer LSTM.
The relative-position feature extractor first computes the relative 3D displacement of the VUE and the BS at each time slot. Since the candidate links evolve continuously with VUE mobility and exhibit strong temporal correlations, the LSTM layer captures the temporal evolution of the VUE-BS relative position from the historical sequence. At the $t$-th time slot, the relative-position feature extractor outputs the relative position feature vector by distance, azimuth, and elevation:
$
\boldsymbol{\gamma}_{b,t}
=
[r_{b,t},\,\theta_{b,t},\,\varphi_{b,t}]^{\top}
\in\mathbb{R}^{3},
$
where
\begin{align}
r_{b,t}
&= \sqrt{\Delta x_{b,t}^{2}+\Delta y_{b,t}^{2}+\Delta z_{b,t}^{2}},
   & \\
\theta_{b,t}
&= \operatorname{atan2}\bigl(\Delta y_{b,t},-\Delta x_{b,t} \bigr)
   \in(-\pi,\pi],
   &  \\
\varphi_{b,t}
&= \arcsin\Bigl(\Delta z_{b,t}/r_{b,t}\Bigr)
   \in[0,\pi/2].
\end{align}
The terms $\Delta x_{b,t}=(m^{\rm BS}_{b,t}-\frac{L}{2})\frac{D}{L}$, $\Delta y_{b,t}=(n^{\rm BS}_{b,t}-\frac{L}{2})\frac{D}{L}$, and $\Delta z_{b,t}=h^{\rm BS}_{b}-h_{\rm c}$ denote the relative displacements between the UE and $b$-th BS. The MLP encoder with parameters $\boldsymbol{\theta}^{\mathrm{geo}}$ maps this low-dimensional feature $\boldsymbol{\gamma}_{b,t}$ to a latent space as $\mathbf{g}_{b,t}\in\mathbb{R}^{d_h}$, which is expressed as $\mathbf{g}_{b,t}=f^{\mathrm{geo}}(\boldsymbol{\gamma}_{b,t};\boldsymbol{\theta}^{\mathrm{geo}})$.
For a historical window of length $T$, the geometry embeddings $\mathbf{G}_{b,t}=[\mathbf{g}_{b,t-T+1},\ldots,\mathbf{g}_{b,t}]
\in\mathbb{R}^{T\times d_h}$ serve as the input of the LSTM to capture the dynamic evolution of the relative position features. The output of the LSTM is its hidden state at the last time step, given by $\mathbf{h}_{b,t}
=
f^{\mathrm{L}}\!\left(
\mathbf{G}_{b,t};
\boldsymbol{\theta}^{\mathrm{L}}
\right)\in\mathbb{R}^{d_s}
$, with dimension of $d_s$. 

\subsection{Prediction Heads and Training Objective}

The input to the prediction head is the concatenated representation of the temporal embedding, the LoS feature embedding, and the height-map context embedding, given by
\begin{equation}
\mathbf{z}_{b,t}
=
\bigl[
\mathbf{h}_{b,t};
\mathbf{m}_{b,t};
\mathbf{c}_{t}
\bigr].
\label{eq:prediction_input}
\end{equation}
Next, four task-specific MLP heads produce the next-slot LoS $\tilde{p}^{\mathrm{LoS}}_{b,t+1}$, BS and VUE beam probability $\hat{\boldsymbol{\pi}}^{\mathrm{tx}}_{b,t+1}$ and $\hat{\boldsymbol{\pi}}^{\mathrm{rx}}_{b,t+1}$, and the optimal-beam transmission rate $\hat{R}^\star_{b,t+1}$, respectively, with corresponding parameters $\boldsymbol{\theta}^{\mathrm{LoS}}$, $\boldsymbol{\theta}^{\mathrm{tx}}$, $\boldsymbol{\theta}^{\mathrm{rx}}$, and $\boldsymbol{\theta}^{\rm Rate}$, which are given by
\begin{align}
\hat{p}^{\mathrm{LoS}}_{b,t+1}
&\!= \mathbb{I}(\underbrace{
\operatorname{sigmoid}(f^{\mathrm{LoS}}\!(\mathbf{z}_{b,t};\boldsymbol{\theta}^{\mathrm{LoS}}) )}_{\tilde{p}^{\mathrm{LoS}}_{b,t+1}}>0.5) \in \{0,1\},\\
\hat{\boldsymbol{\pi}}^{\mathrm{\kappa}}_{b,t+1}
&\!=\!\operatorname{softmax}\!\left(f^{\mathrm{\kappa}}\!\left(\mathbf{z}_{b,t};\boldsymbol{\theta}^{\mathrm{\kappa}}\right)\right)\in [0,1]^{C_{\mathrm{\kappa}}},\label{eq:tx_beam_distribution}\\
\hat{R}^{\star}_{b,t+1}
&\!=\!\operatorname{softplus}\left(f^{R}\!\left(\mathbf{z}_{b,t};\boldsymbol{\theta}^{\mathrm{Rate}}\right)\right)\in\mathbb{R}_{+},
\end{align}
where $\kappa\in\{\mathrm{tx},\mathrm{rx}\}$ and $\tilde{p}^{\mathrm{LoS}}_{b,t+1}$ is LoS probability.
The BS and VUE beam rankings are obtained by sorting the beam probabilities in descending order and keeping the first $K_{\mathrm{beam}}$ indices:
\begin{equation}
\hat{\mathcal{R}}^{\kappa}_{b,t+1}
\!=\!
\operatorname*{argsort_{\downarrow}}
\!\left(\hat{\boldsymbol{\pi}}^{\kappa}_{b,t+1}\right)[1:K^{\kappa}_{\mathrm{beam}}],
\,
\forall \kappa\in\{\mathrm{tx},\mathrm{rx}\}.
\label{eq:beam_ranking}
\end{equation}


Given the ground-truth labels ${p}^{\mathrm{LoS}}_{b,t+1}$, ${R}^\star_{b,t+1}$, ${\boldsymbol{\pi}}^{\mathrm{tx}}_{b,t+1}$ and ${\boldsymbol{\pi}}^{\mathrm{rx}}_{b,t+1}$ at the $b$-th candidate BS at time slot $t+1$, MMST is trained with a multi-task objective based on homoscedastic uncertainty weighting~\cite{cipollaMultitaskLearningUsing2018a}. 
Let $\ell_{\mathrm{LoS}}=\operatorname{BCE}(\tilde p^{\mathrm{LoS}}_{b,t+1},p^{\mathrm{LoS}}_{b,t+1})$, $\ell_\mathrm{R}=\operatorname{Smooth}_{\ell_1}(\hat R^\star_{b,t+1},R^\star_{b,t+1})$, $\ell_{\kappa}=\operatorname{CE}(\hat{\boldsymbol{\pi}}^{\kappa}_{b,t+1},\boldsymbol{\pi}^{\kappa}_{b,t+1}), \forall \kappa\in\{\mathrm{tx},\mathrm{rx}\}$,  
denote the binary cross-entropy, smooth-$\ell_1$~\cite{girshick2015fastrcnn}, and cross-entropy loss.
Then, the loss function for MMST is given by
\begin{equation}\label{eq:mmst_loss}
\mathcal{L}(\bm{\theta}^{\mathrm{MMST}})
=\sum_{\chi\in\{\mathrm{LoS},\mathrm{tx},\mathrm{rx},\mathrm{R}\}}
\left(\frac{w_\chi \ell_\chi}{2\lambda_\chi^2}+\frac{1}{2}\log\lambda_\chi^2\right)
\end{equation}
where ${\bm \theta}^{\mathrm{MMST}} \triangleq \{{\bm \theta}^{\mathrm{CNN}},{\bm \theta}^{\mathrm{sp}},{\bm \theta}^{\mathrm{geo}},{\bm \theta}^{\mathrm{L}},{\bm \theta}^{\mathrm{LoS}},{\bm \theta}^{\mathrm{tx}},{\bm \theta}^{\mathrm{rx}},{\bm \theta}^{\mathrm{Rate}}\}$, $w_{\mathrm{LoS}}=1$, $w_{\mathrm{tx}}=w_{\mathrm{rx}}=w_\mathrm{R}=p^{\mathrm{LoS}}_{b,t+1}$, and $\lambda_\chi$ is the learnable uncertainty parameter for the corresponding task. $\lambda_\chi$ is initialized to $1$ and dynamically adjusted during training. The LoS label $p^{\mathrm{LoS}}_{b,t+1}$ gates the beam ranking and data rate because the optimal beam and transmission-rate labels are only meaningful for valid LoS links.


\section{Performance Evaluation}

\subsection{Dataset}
\label{sec:data_generation}
The dataset has two parts: the visual dataset for training the VBHE, and the communication channel dataset for training the MMST. The visual data set has a total of 17K samples in  18 different real-world regions of 4 cities covering around $88.0~\mathrm{km}^2$ in New South Wales, Australia, named the Hills Shire, Blacktown, Parramatta, and Cumberland, consisting of street-view panoramic images, satellite images, and their building footprint and height maps. The street-view panoramic images and satellite images are collected from Google Maps~\cite{maps-static-api}. These regions include the city center, the commercial region, the residential region, and the industrial region. The building footprint and height maps are obtained from OpenStreetMap~\cite{osm2026}. 
The communication channel data set with 290K samples is generated by Sionna~{\cite{hoydis2023sionna}}, which utilizes ray-tracing techniques to simulate the LoS indicators, optimal-beam data rates, and beam indices based on the pre-defined VUE trajectories and candidate-BS coordinates in the same wireless propagation environment in these 18 regions. Due to space limitations, more details of the data set generation process are available in our GitHub project repository.


\subsection{Experiment Settings}
The system parameters for the configuration of the BS and VUE, and the channel conditions are summarized in Table~\ref{tab:sim_params}.
\begin{table}[t]
\centering
\footnotesize
\setlength{\tabcolsep}{5pt}
\caption{Simulation Parameters}
\label{tab:sim_params}
\vspace{-0.2cm}
\begin{tabular}{m{6.2cm} m{1.8cm}}
\toprule
\textbf{Parameter} & \textbf{Value} \\
\midrule
\rowcolor[HTML]{EFEFEF} 
Number of BS antennas, $N\!=\!N_xN_y$ & $12\times 8$\\
Number of VUE antennas, $M\!=\!M_xM_y$ & $4\times 4$ \\
\rowcolor[HTML]{EFEFEF} 
 Beam codebook sizes of BS/VUE, $C_{\mathrm{tx}}$; $ C_{\mathrm{rx}}$ & $96$; $16$ \\
Beam-candidate size of BS/VUE, $K^{\rm tx}_{\mathrm{beam}}$; $ K^{\rm rx}_{\mathrm{beam}}$ & $10$; $5$\\
\rowcolor[HTML]{EFEFEF} 
BS transmission power, $P_{\rm tot}$ & 30 dBm \\
Carrier frequency, $f_c$  & 28 GHz \\
\rowcolor[HTML]{EFEFEF} 
Subcarrier spacing & 120 kHz \\
Number of subcarriers, $K$ & 1584 \\
\rowcolor[HTML]{EFEFEF} 
Bandwidth, & 200 MHz \\
Noise power density & $-174$ dBm/Hz \\
\rowcolor[HTML]{EFEFEF} 
Slot duration, $\tau_{\mathrm{s}}$ & $33.33$ ms \\
Beam-sweep unit time $\tau_{\mathrm{u}}$ & $0.03568$ ms \\

\rowcolor[HTML]{EFEFEF} 
Handover interruption, $\tau_{\mathrm{h}}$ & $20$ ms \\

BS density (100\%) & $0.0019\,\mathrm{m}^{-2}$ \\
\rowcolor[HTML]{EFEFEF} 
Maximum paths per receiver, $L_{\rm p}$ & 32 \\

{Velocity} of VUE & 43.2 km/h \\
\rowcolor[HTML]{EFEFEF} 
Average GPS error & 0.4363 m \\
\bottomrule
\end{tabular}
\vspace{-0.5cm}
\end{table}

\begin{table*}[t]
\centering
\caption{Comparison of VBHE with cross-view transformation baselines under different evaluation protocols.}
\label{tab:VBHE_baseline_comparison}
\vspace{-0.3cm}
\footnotesize
\setlength{\tabcolsep}{11.5pt}

\begin{threeparttable}
\begin{tabular*}{\textwidth}{lcccccccc}
\toprule
\multirow{3}{*}{\textbf{Method}}
& \multicolumn{4}{c}{\textbf{North-West Sydney (Hills Shire, Blacktown)}}
& \multicolumn{4}{c}{\textbf{Central-West Sydney (Parramatta, Cumberland)}} \\
\cmidrule(lr){2-5} \cmidrule(lr){6-9}
& \multicolumn{2}{c}{\textbf{In-region}}
& \multicolumn{2}{c}{\textbf{Cross-region}}
& \multicolumn{2}{c}{\textbf{In-region}}
& \multicolumn{2}{c}{\textbf{Cross-region}} \\
\cmidrule(lr){2-3} \cmidrule(lr){4-5}
\cmidrule(lr){6-7} \cmidrule(lr){8-9}
& \textbf{mIoU} $\uparrow$ & \textbf{MAE} $\downarrow$
& \textbf{mIoU} $\uparrow$ & \textbf{MAE} $\downarrow$
& \textbf{mIoU} $\uparrow$ & \textbf{MAE} $\downarrow$
& \textbf{mIoU} $\uparrow$ & \textbf{MAE} $\downarrow$ \\
\midrule

\rowcolor[HTML]{EFEFEF} 
GP~\cite{shiBoosting3DoFGroundtoSatellite2023}
& 77.18 & 1.53
& 54.68 & 5.57
& 79.22 & 2.17
& 67.62 & 2.23 \\

ST~\cite{wangFineGrainedCrossViewGeoLocalization}
& 82.88 & 1.74
& 72.93 & 2.43
& 81.78 & 1.84
& 65.72 & 2.19 \\

\rowcolor[HTML]{EFEFEF} 
\textbf{VBHE}
& \textbf{84.49} & \textbf{1.33}
& \textbf{75.93} & \textbf{2.18}
& \textbf{83.88} & \textbf{1.64}
& \textbf{68.63} & \textbf{1.90} \\ \hline

VBHE w/o DPT
& 82.89 & 1.44
& 73.62 & 2.31
& 79.76 &  2.70                         
& 63.32 & 2.98 \\

\rowcolor[HTML]{EFEFEF} 
VBHE w/o Roof Projection
& 80.85 & 1.53
& 74.86 & 2.41
& 80.41 & 2.04
& 67.14 & 2.16 \\

VBHE w/o Ground Projection
& 81.24 & 1.52
& 74.53 & 2.23
& 77.50 & 2.93
& 66.80 & 3.09 \\

\bottomrule
\end{tabular*}
\end{threeparttable}
\vspace{-0.5cm}
\end{table*}

\subsubsection{Hyper-Parameters of VBHE}
$L=256$, $H=416$, $W=832$, $C=3$, $D=79.36$, $D_r=16$, $D_b=32$, $h_{\rm c} = 1.5$~m, $H'=52$, and $W'=104$.
For training, the weights of the mask and height losses are set to $\lambda_m=1$ and $\lambda_h=3.5$. We use \textit{Adam} as the optimizer with an initial learning rate of $6\times10^{-5}$ and a weight decay of $1\times10^{-2}$. A step learning-rate schedule is adopted, where the learning rate is decayed by a factor of $0.7$ every 10 epochs. The model is trained for 50 epochs with a batch size of 16. 
\subsubsection{Hyper-Parameters of MMST} $T=5$, $L=256$, $d_s=256$, $d_h=128$, $d_m=128$, $Q=64$, and $R=3$. 
The model is trained for 80 epochs with a batch size of 512. During training, we use \textit{Adam} as the optimizer with an initial learning rate of $5\times10^{-4}$.
\subsubsection{Performance Metrics}
For VBHE, we use the mean intersection over union (mIoU) as the evaluation metric for building-mask segmentation performance~\cite{everinghamPascalVisualObject2015}. For height-map estimation, we use mean absolute error (MAE) between the predicted and the ground-truth heights within the set of all building-cell indices in $\mathbf M$.
For MMST, we evaluate the performance of binary LoS prediction by using the accuracy, precision, recall, and F1 score~\cite{powersEvaluationPrecisionRecall2011}. For data-rate prediction, we use the MAE between the  $\hat{R}^\star_{b,t+1}$ and ${R}^\star_{b,t+1}$ as the evaluation metric. 
For beam prediction, the model outputs a set of $K^{\kappa}_{\mathrm{beam}}, \forall \kappa\in\{\mathrm{tx,rx}\}$ candidate beams for each sample. A prediction is considered correct if the ground-truth beam index is included in this candidate set and incorrect otherwise. The top-\(K^{\kappa}_{\mathrm{beam}}\) prediction accuracy is then defined as the proportion of correctly predicted samples among all evaluated samples.

\subsection{Baselines}
The proposed framework comprises two stages: VBHE and MMST, which are evaluated separately against different task-specific baselines.
\subsubsection{Baselines for VBHE}
\textbf{\textit{GP}}~\cite{shiBoosting3DoFGroundtoSatellite2023}:
The geometry projection (GP) method uses deterministic camera geometry and ground-plane homography to project ground-view image features into an overhead-view representation. 
The projected features therefore encode ground-plane visual cues in the BEV coordinate system, while scene structures above the ground plane are handled only through the geometric projection assumption. \textbf{\textit{ST}}~\cite{wangFineGrainedCrossViewGeoLocalization}:
The spherical transform (ST) method maps spherical coordinates in the ground-camera system to normalized equirectangular panoramic coordinates using inverse trigonometric functions.

\subsubsection{Baselines for MMST}
\textbf{\textit{GBPN}}~\cite{liOutofBandModalitySynergyBased2026a}:
 In its design, a CNN module extracts spatial features from each encoding map, a GRU module captures temporal evolution across consecutive frames, and task-specific prediction heads estimate the beamforming gain and the optimal beam. \textbf{ \textit{HMCN}}~\cite{fengEnvironmentSensingAidedBeam2025}:
The method extracts dynamic features from semantic maps using 2D CNNs, extracts static environmental features from point-cloud-derived pseudo image sequences using 3D CNNs, and encodes user-location information through fully connected layers. The dynamic, static, and user-location features are then fused and fed into a temporal predictor to estimate future beam indices. \textbf{ \textit{TORP}}~\cite{ahnSensingComputerVisionAided2024}:
The method uses the Transformer encoder by constructing a sequence of visual environmental and geometric tokens to infer the link data rate. The predicted results are then used for proactive BS selection. \textbf{\textit{5G NR Reactive Baselines}}:
We follow the NR Event A3 condition~\cite{3gppTS38331}, where a neighboring BS becomes sufficiently better than the serving BS, triggering the handover decision. The beam-sweeping overhead is 
$    \tau_{\mathrm{m}}
    =
    \left(C_{\mathrm{tx}}+C_{\mathrm{rx}}\right)\tau_{\mathrm{u}}$.
\textbf{\textit{Reactive HBS:}} Reactive HBS has the same NR Event A3 condition handover strategy, while the exhaustive sweeping is replaced with hierarchical beam search~\cite{xiaoHierarchicalCodebookDesign2016} where the number of searched beams is set to 
$
    2\left\lceil
    \log_2
    \left(
    C_{\mathrm{tx}}C_{\mathrm{rx}}
    \right)
    \right\rceil$.

\subsection{Evaluation of VBHE}

We first evaluate the building segmentation and height-prediction performance of the proposed VBHE module and compare it with the baselines across the 18 regions. We report results for two target areas: North-West Sydney, comprising Hills Shire and Blacktown, and Central-West Sydney, comprising Parramatta and Cumberland. For each target area, we consider both in-region and cross-region evaluation. In the in-region setting, samples within the target area are randomly partitioned into training, validation, and test sets at a ratio of 0.8:0.1:0.1. In the cross-region setting, each model is tested in the opposite region. In other words, the test area is completely unseen by the model during training phases. This setting, therefore, measures the generalization of VBHE in the unseen regions. 

\begin{figure}[t]
\centerline{\includegraphics[width=0.9\linewidth]{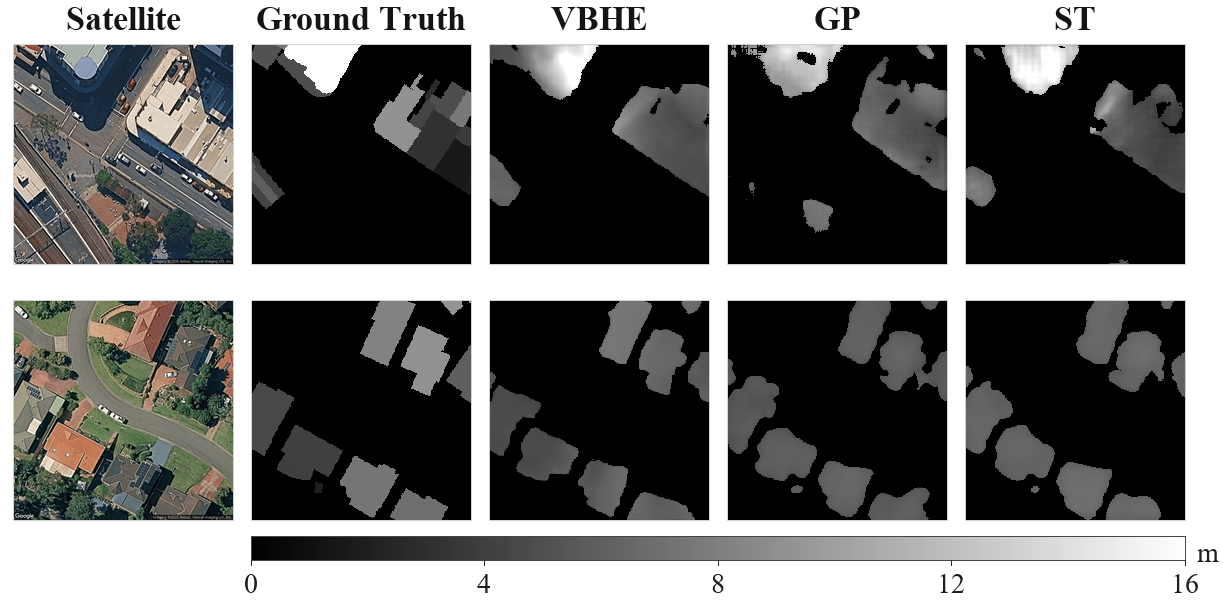}}
\vspace{-0.4cm}
\caption{Visualization of the height map of different methods.}
\label{vbhe-sample}
\vspace{-0.5cm}
\end{figure}
Fig.~\ref{vbhe-sample} illustrates the estimated height map of VBHE and the GP and ST baselines in urban and suburban areas. The results demonstrate that VBHE produces more accurate building segmentation and height prediction than the baselines, which are crucial for accurate environmental reconstruction. In contrast, GP tends to produce distorted building shapes due to its reliance on ground projection, while ST tends to over-smooth building boundaries in this complex scene.

Table~\ref{tab:VBHE_baseline_comparison} illustrates the mIoU and MAE results for the two target areas under both in-region and cross-region settings. The results show that VBHE consistently outperforms the cross-view transformation baselines, achieving higher mIoU and lower MAE in both in-region and cross-region evaluations. 
For the cross-region evaluation, although the performance of all methods decreases, VBHE still outperforms the baselines, demonstrating its ability to generalize to unseen areas. The results highlight the effectiveness of the proposed VBHE module in capturing building segmentation and height prediction from street-view and satellite images, which is crucial for accurate environmental reconstruction and subsequent communication prediction tasks. Moreover, the ablation results show that removing the DPT, Ground Projection, or Rooftop Projection consistently reduces the mIoU and increases the height MAE, validating the effectiveness of each component of VBHE  towards overall performance. 

We note that, to emulate a realistic deployment scenario, all subsequent evaluations are conducted under the cross-region setting, with North-West Sydney serving as a geographically unseen test region.
\begin{table*}[t]
\centering
\caption{Multi-task prediction performance on the common evaluation set.}
\label{tab:block_test_main}
\vspace{-0.3cm}
\begingroup
\footnotesize
\setlength{\tabcolsep}{2.6pt}

\begin{threeparttable}
\begin{tabular*}{\textwidth}{l cccc c ccc}
\toprule
\multirow{2}{*}{\textbf{Method}}
& \multicolumn{4}{c}{\textbf{LoS Classification} $\uparrow$}
& \multicolumn{1}{c}{\textbf{Rate} $\downarrow$}
& \multicolumn{3}{c}{\textbf{Top-1 Beam Prediction} $\uparrow$} \\
\cmidrule(lr){2-5} \cmidrule(lr){6-6} \cmidrule(lr){7-9}
& \textbf{Acc. (\%)} & \textbf{Prec. (\%)} & \textbf{Rec. (\%)} & \textbf{F1 (\%)}
& \textbf{MAE (bps/Hz)}
& \textbf{BS Acc.(\%)} & \textbf{VUE Acc.(\%)} & \textbf{Joint Acc.(\%)} \\
\midrule

GTHM-MMST
& 98.0 & 98.0 & 99.2 & 98.6 & 0.589 & 98.4 & 99.6 & 98.4 \\
\midrule

\textbf{VBHE-MMST}
& \textbf{91.4} & \textbf{92.9} & \textbf{94.7} & \textbf{93.8} & \textbf{0.638} & \textbf{90.8} & \textbf{97.2} & \textbf{89.0} \\

\rowcolor[HTML]{EFEFEF} 
VBHE-GBPN~\cite{liOutofBandModalitySynergyBased2026a}
& 81.8 & 87.0 & 86.5 & 86.7 & 0.786 & 89.2 & 96.2 & 88.2 \\
VBHE-HMCN~\cite{fengEnvironmentSensingAidedBeam2025}
& 72.6 & 82.7 & 76.1 & 79.2 & 0.869 & 85.7 & 95.0 & 83.1 \\

\rowcolor[HTML]{EFEFEF} 
VBHE-TORP~\cite{ahnSensingComputerVisionAided2024}
& 67.9 & 84.0 & 65.9 & 73.9 & 0.757 & 88.8 & 95.0 & 85.9 \\
\midrule
VBHE-MMST w/o Height-map Context Feature
& 80.2 & 84.9 & 86.5 & 85.7 & 0.745 & 89.3 & 95.8 & 86.6 \\

\rowcolor[HTML]{EFEFEF} 
VBHE-MMST w/o LoS Feature
& 66.6 & 73.8 & 79.8 & 76.7 & 1.049 & 74.6 & 86.6 & 68.1 \\
VBHE-MMST w/o Relative-Position Feature
& 77.6 & 83.9 & 83.4 & 83.6 & 1.381 & 68.6 & 65.0 & 44.2 \\

\rowcolor[HTML]{EFEFEF} 
VBHE-MMST w/o Height Map
& 61.0 & 69.3 & 77.9 & 73.3 & 1.227 & 72.5 & 83.3 & 64.9 \\
\bottomrule
\end{tabular*}

\end{threeparttable}
\endgroup
\vspace{-0.5cm}
\end{table*}
\begin{figure}[t]
\centerline{\includegraphics[width=\linewidth]{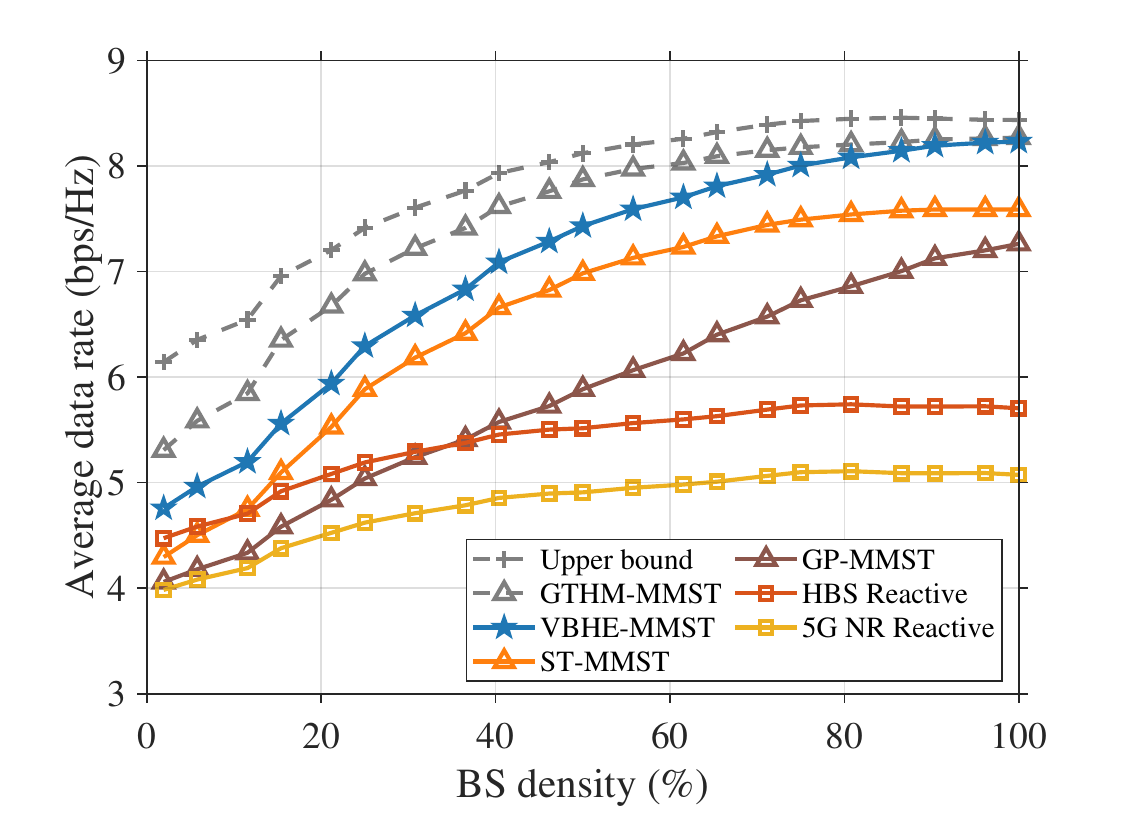}}
\vspace{-0.2cm}
\caption{{Comparison of VBHE with baselines under different BS densities.}}
\label{density-baseline-mmst}
\vspace{-0.4cm}
\end{figure}

\begin{figure}[t]
\centerline{\includegraphics[width=\linewidth]{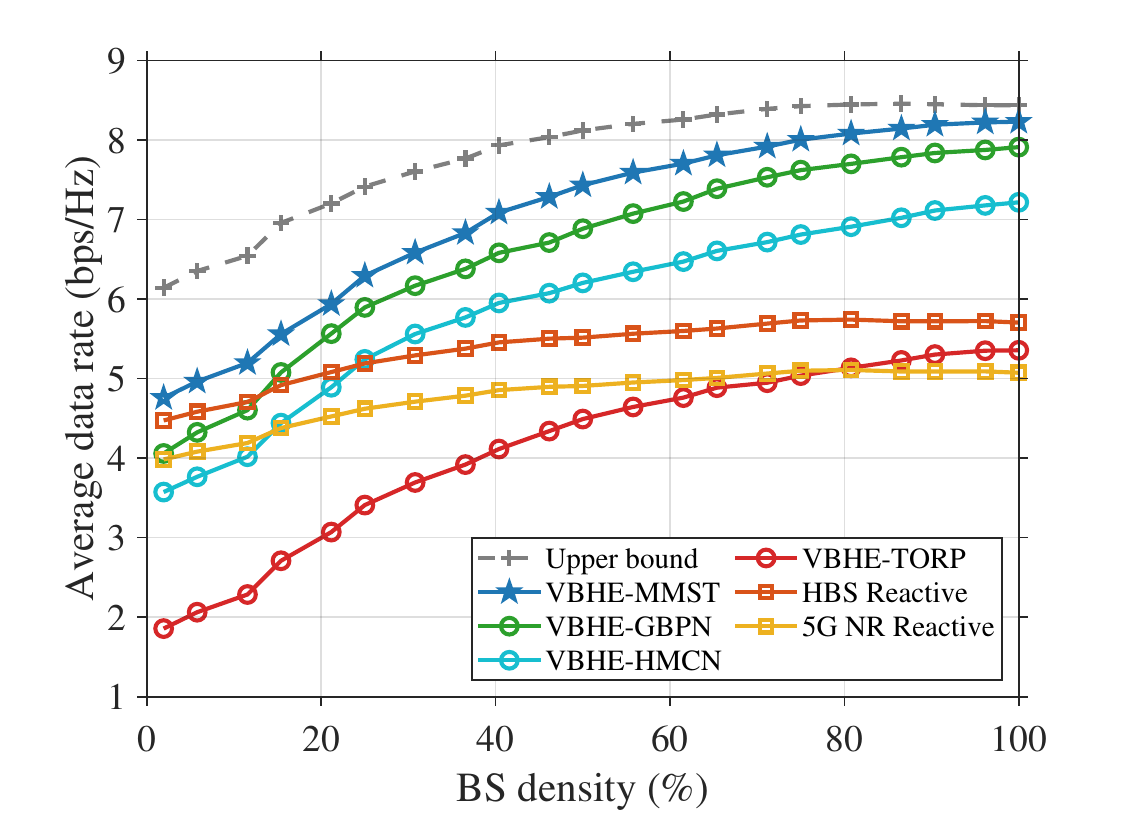}}
\vspace{-0.1cm}
\caption{Comparison of MMST with baselines under different BS densities.}
\label{density-baseline-vbhe}
\vspace{-0.5cm}
\end{figure}

\subsection{Evaluation of MMST}
\label{subsec:mmst_evaluation}
We further evaluate the performance of the proposed MMST model in three aspects: LoS classification, predicted data rate, and beam index selection of BS and VUE for the next time slot. Similar to the VBHE, the training, validation, and testing datasets are constructed from trajectories in geographically disjoint regions with partitioning of 0.8:0.1:0.1, ensuring that both VBHE and MMST are evaluated in the same geographically unseen area. 

Table~\ref{tab:block_test_main} reports the prediction performance of the proposed MMST model and the baselines. All methods are evaluated with the same VBHE-estimated height map as the environmental input. GTHM-MMST uses the ground-truth height map thus it provides an upper bound for assessing the impact of height-map estimation errors on MMST. VBHE-MMST achieves the best performance across all three prediction tasks. For LoS classification, VBHE-MMST achieves an accuracy of $91.4\%$, whereas GBPN, HMCN, and TORP achieve $81.8\%$, $72.6\%$, and $67.9\%$, respectively. For the MAE of the rate prediction, VBHE-MMST obtains the lowest MAE of $0.638~\mathrm{bps/Hz}$, compared with $0.786~\mathrm{bps/Hz}$ for GBPN, $0.869~\mathrm{bps/Hz}$ for HMCN, and $0.757~\mathrm{bps/Hz}$ for TORP. VBHE-MMST also achieves the highest Top-1 ($K_{\mathrm{beam}}^{\kappa}=1$) beam prediction accuracy compared with GBPN, HMCN,  and TORP. For example, the MMST achieves $90.8\%$ BS beam prediction accuracy, while other baselines achieve $89.2\%$, $85.7\%$, and $88.8\%$, respectively. These results demonstrate that the proposed MMST model effectively leverages the visual height-map representation to improve link-level prediction.

The ablation results in Table~\ref{tab:block_test_main} show that removing any MMST component degrades the overall performance, validating the complementary contributions of the height-map context feature, LoS features, relative-position features, and height-map input to LoS, rate, and beam prediction.

\subsection{Evaluation of the Entire Framework}



\subsubsection{Performance under Varying BS Densities}
We evaluate the proposed framework under varying BS densities, ranging from sparse to dense deployments. Fig.~\ref{density-baseline-mmst} first evaluates the impact of the height-map estimation module by applying the same MMST predictor to the height maps produced by GTHM, VBHE, ST, and GP, respectively. We note that the upper bound is computed in Sionna using the ground-truth building environment with ground-truth link prediction and therefore serves as an ideal reference. Across the entire density range, VBHE-MMST achieves a throughput close to that of GTHM-MMST and consistently outperforms other methods. For example, at a BS density of $100\% (0.0019\,\mathrm{m}^{-2})$ and velocity of $32.4~\mathrm{km/h}$, VBHE-MMST achieves an average data rate of around $8.2~\mathrm{bps/Hz}$, while ST-MMST and GP-MMST achieve $7.57~\mathrm{bps/Hz}$  and $7.36~\mathrm{bps/Hz}$, respectively. These results demonstrate the effectiveness of VBHE in accurately estimating the building geometry and blockage information required for downstream communication decisions under varying BS densities. In contrast, reactive 5G~NR exhibits limited throughput improvement as the BS density increases. Fig.~\ref{density-baseline-vbhe} compares different link-prediction models using the same VBHE-estimated environmental representation. Similarly, VBHE-MMST consistently achieves the highest throughput among all practical learning-based methods over the full range of BS densities, including VBHE-HMCN, VBHE-GBPN, and VBHE-TORP. 

\subsubsection{Performance under Different VUE Velocities}

\begin{figure}[t]
\centerline{\includegraphics[width=0.97\linewidth]{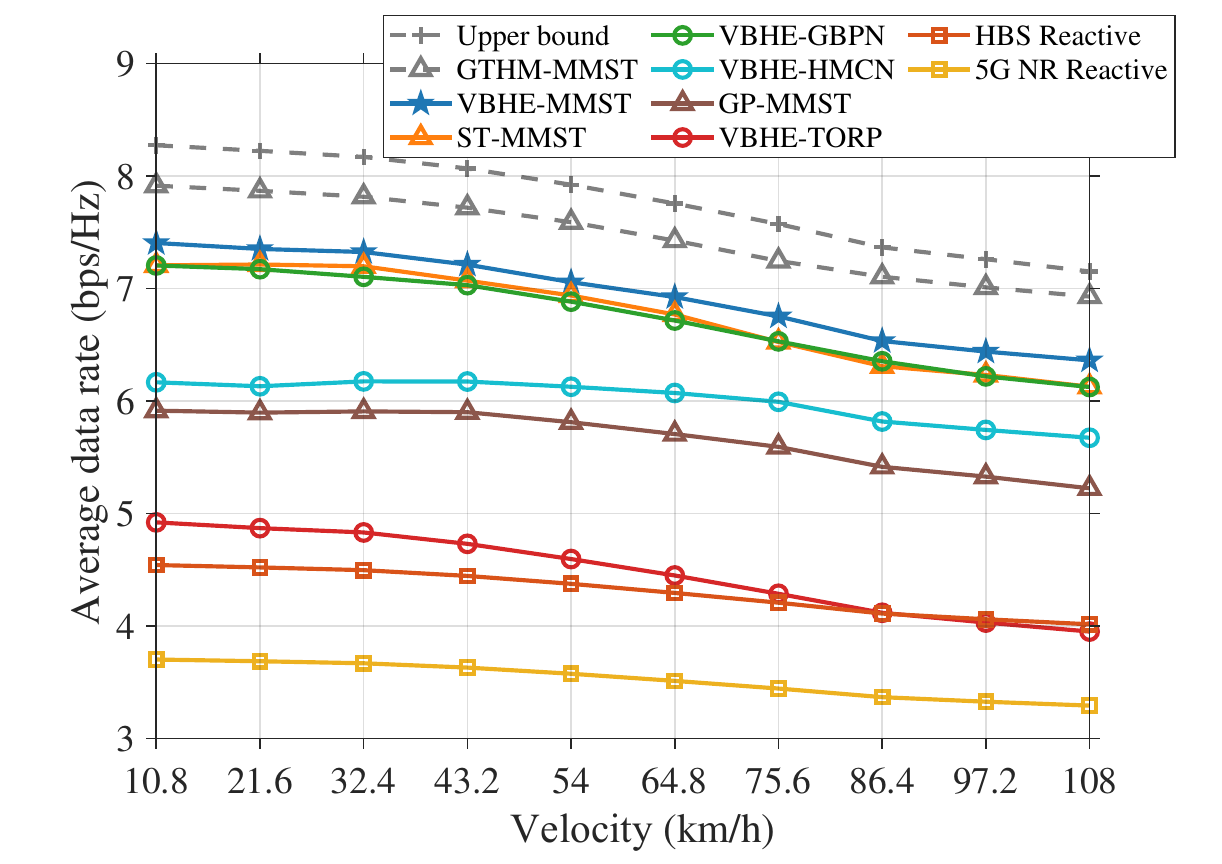}}
\vspace{-0.2cm}
\caption{Average data rate versus VUE velocity.}
\label{speed}
\vspace{-0.3cm}
\end{figure}
We evaluate the proposed framework under different VUE velocities, ranging from $10.8$ to $108~\mathrm{km/h}$. The evaluation is conducted under the same BS density of $80\%$. As shown in Fig.~\ref{speed}, the average transmission data rate decreases as the VUE velocity increases. This degradation is mainly attributed to two factors. First, a higher velocity causes rapid channel changes, making the predicted link information more likely to become outdated and reducing the data rate in the same time slot. Second, a faster-moving VUE crosses BS coverage and blockage boundaries more frequently, leading to more frequent BS association and handover. This overhead reduces the time available for data transmission. Despite these challenges, VBHE-MMST consistently achieves the highest data rate among all practical schemes across the evaluated velocity range, demonstrating its robustness to mobility-induced channel variation and link dynamics.

\subsubsection{Performance under Different BS Antenna Numbers}
\begin{figure}[t]
\centerline{\includegraphics[width=0.95\linewidth]{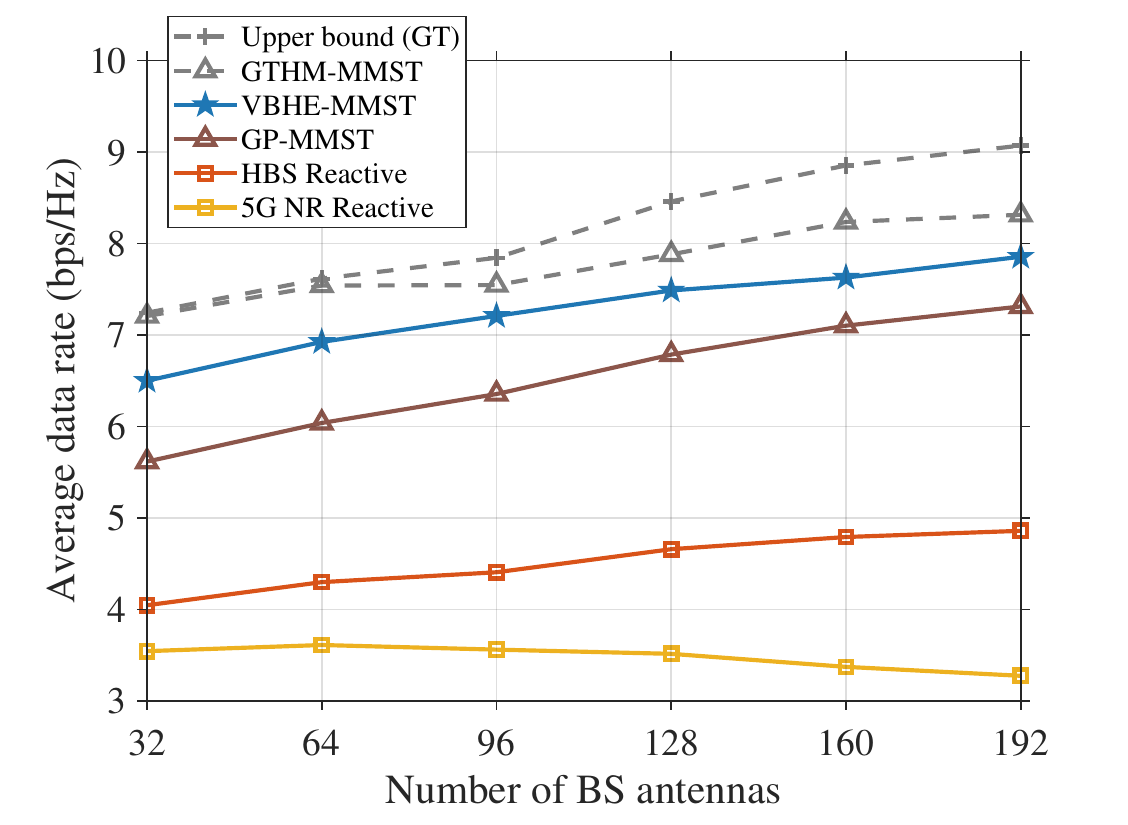}}
\vspace{-0.2cm}
\caption{Average data rate versus the BS antenna number.}
\label{beam_number}
\vspace{-0.5cm}
\end{figure}
We evaluate the proposed framework with different BS antenna array sizes, ranging from $N=32$ to $192$ antennas. Fig.~\ref{beam_number} compares VBHE-MMST with the baselines and the upper bound. The results show that the average transmission data rate increases as the number of BS antennas increases. This is because a larger antenna array provides higher beamforming gains under the same transmission power, which improves the data rate during the data transmission period. Among all practical methods, VBHE-MMST consistently achieves the highest data rate across the evaluated antenna numbers. This demonstrates the scalability and effectiveness of the proposed framework in different BS antenna configurations. We note that the data rate of 5G~NR decreases as the number of BS antennas increases. Although a larger antenna array provides a higher beamforming gain, it also enlarges the transmit beam codebook size \(C_{\mathrm{tx}}\). Consequently, the exhaustive beam-search procedure of 5G~NR must evaluate more candidate beams, resulting in greater beam-training overhead and a shorter duration for data transmission. In this case, the increased search overhead outweighs the additional beamforming gain, leading to a reduction in the average data rate.

\subsubsection{Computation Overhead}
\label{subsubsec:computation_overhead}
Online inference complexity of VBHE and MMST is measured by the number of on-device parameters and the floating-point operations (FLOPs) required to process one sample. VBHE requires 9.95~M trainable parameters and 88.098~GFLOPs per sample, whereas MMST requires $5.94$~M deployed model parameters and $0.152$~GFLOPs per sample. We also evaluate the inference latency of each sample on a desktop with the NVIDIA GeForce RTX 4090 GPU. Specifically, the inference latency is averaged over $10{,}000$ samples. The inference latency of VBHE and MMST is 9.10~ms and 0.71~ms, respectively, which is significantly lower than the duration of each time slot, demonstrating its suitability for V2I link prediction in practical deployment scenarios.

\section{Conclusion}

\ifCLASSOPTIONcaptionsoff
  \newpage
\fi
\bibliographystyle{IEEEtran}

This paper proposed an environmental-sensing-aided framework for proactive BS selection and beam prediction in mmWave V2I communication systems. Specifically, VBHE first estimates the building height map from onboard panoramic street-view images and the preloaded satellite map. Then, MMST combines the estimated environmental features with historical mobility information to predict next-slot link states, rates, and beam rankings. Extensive experiments in geographically unseen urban regions show that the proposed framework outperforms the evaluated deployable baselines, demonstrating that the height maps can provide useful geometric information for proactive V2I communications.

\bibliography{bibtex/bib/IEEEabrv,scriptbib}
\end{document}